\documentclass{elsarticle}
\journal{}

\usepackage{amssymb}

\usepackage{float}

\begin{document}

\begin{frontmatter}



\title{Improved $b$-jet Energy Correction for $H \to b\bar{b}$ Searches at CDF}


\author[label1]{T. Aaltonen}
\author[label2]{A. Buzatu}
\author[label3]{B. Kilminster}
\author[label4]{Y. Nagai}
\author[label5]{W. Yao}

\address[label1]{Helsinki Institute of Physics and University of Helsinki, Finland, 00014}
\address[label2]{McGilll University, Rutherford Physics Building, Montreal, Canada, H3A2T8}
\address[label3]{Fermi National Accelerator Laboratory, Batavia, Illinois, 60510}
\address[label4]{Graduate School of Pure and Applied Sciences, University of Tsukuba, Japan, 305-8571}
\address[label5]{Ernest Orlando Lawrence Berkeley National Laboratory, Berkeley, California 94720}


\begin{abstract}
We present a method for improving the $b$-jet energy resolution in order to improve 
the signal sensitivity in searches for particles decaying to a $b$ quark and anti-$b$ quark.  
A correction function is computed for individual jets, 
which combines information from the secondary vertex tagger, the offline tracking 
and standard calorimeter-based jet-energy reconstruction algorithm 
in order to provide a more accurate measurement of the true 
$b$-quark energy. We apply the correction to Monte-Carlo-simulated jets in the process  
$WH \rightarrow \ell \nu b\bar b$ and find an improvement in both the mean and the resolution 
of the $b$-jet energy with respect to the $b$-quark energy.  The correction 
improves the measured Higgs dijet invariant mass resolution from
$\sim$ 15\%(standard jet corrections) to $\sim$ 11\%(improved jet corrections) in the 
Higgs mass range from 100 GeV/$c^2$ - 150 GeV/$c^2$. Using the corrected $b$-jet 
energies instead of the standard calorimeter-based $b$-jet energies results in a 
$\sim$ 9\% improvement in the expected sensitivity for Higgs boson production cross section
in the most sensitive search region of the $WH \rightarrow \ell \nu b\bar b$ analysis, which is two tagged jets and
one charged central lepton.
\end{abstract}

\end{frontmatter}

\tableofcontents


\section{Introduction}
\label{intro}

Quarks and gluons, together called partons, are elementary particles produced in high-energy particle physics collisions,
and are identified in particle-detectors by collimated sprays of energetic particles called jets.
The quark four-momenta can be estimated using
energy measurements from calorimeters or charged-particle momenta measurements using tracking detectors.  Heavy quarks called ``$b$´´ quarks
are produced in 2$\rightarrow$2 QCD processes and are especially interesting since they also arise from physics
processes such as top-quark decay, Higgs-boson decay, and signatures of hypothetical particles predicted beyond the standard model,
such as those in supersymmetric theories. These $b$ quarks are different from
light quarks, such as $u$, $d$, $s$ quarks, and gluons, because they can be ``tagged´´ due to the significant distance they traverse from
the primary vertex of the collision before decaying in a secondary vertex with charged particle tracks. For low mass Higgs boson searches,
between 100 GeV/$c^2$ and 135 GeV/$c^2$, in which a Higgs boson is most likely to decay to a $b$ quark and an anti-$b$ quark, the Higgs boson mass can be
reconstructed by calculating the invariant mass of the two jets associated with the identified $b$ quarks.  In $H \to b\bar{b}$ decays,
this ``dijet´´ invariant mass produces a Gaussian-like resonance, which can be distinguished from the smoothly falling exponential-like dijet mass
of the background processes, and is the most effective discriminant to distinguish Higgs bosons from backgrounds. The dijet mass in events with
tagged $b$ jets is the basis for the Tevatron program to search for the Higgs boson in the low mass region. In order to reconstruct and
identify physics processes such as Higgs boson production, the jet-energy resolution defined as the RMS of $(Et_{gen}-Et)/Et$, where $Et$ is the measured
jet energy in the transverse plane and $Et_{gen}$ is the Monte-Carlo generated true value of the particle producing the jet, should be as
reduced as possible. Similarly, $(Et_{gen}-Et)/Et$ should be as close to zero as possible to provide correct $b$-quark energy. Without this
requirement, analyses need special calibration curves to correctly measure the dijet mass.
By improving the jet-energy resolution, the dijet mass resolution also improves, and therefore the sensitivity of the analysis to measure
low production cross sections for new physics processes increases as well. In this article, we describe an algorithm to provide constraints 
on the jet-energy resolution for $b$ quarks by incorporating precision tracking measurements and  secondary vertex information, in addition to standard
calorimeter measurements. This algorithm also allows the 
true $b$-quark energy to be more accurately calculated.  We apply this algorithm to $b$-tagged jets in the CDF search
for the associated production of the Standard Model Higgs boson and $W$ bosons~\cite{Yo_thesis} and determine the improvement
on the sensitivity of this search due to this algorithm. 

\section{Detector}
\label{detector}

\begin{figure}[H]
  \begin{center}
    \includegraphics[width=6.0cm,clip=]{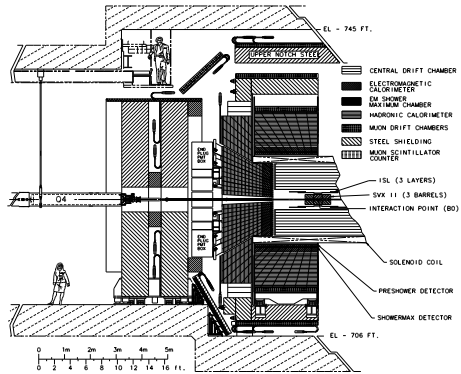}
    \caption{Schematic view of CDF Run II detector.}
    \label{fig_detector}
  \end{center}
\end{figure}

The CDF II detector~\cite{CDF} is a forward-backward and azimuthally symmetric apparatus, shown on Figure~\ref{fig_detector},
for studying $\sqrt{s} = 1.96$ TeV $p\bar{p}$ collisions
at the Fermilab Tevatron.  It consists of a 1.4 T magnetic spectrometer, which provides a magnetic field parallel to the $p$ and $\bar{p}$ beams
enabling charged particle tracking through curvature measurements in the enclosed tracking system.  The tracking inner layers consist of
a 700,000-channel silicon microstrip detector, the Silicon Vertex (SVX) and Intermediate Silicon Layers (ISL), which measures ionization from
charged particles in radial layers from 1.5 to 28 cm, for charged particles $|\eta| <$ 2, where $\eta$ is the pseudo-rapidity defined
as $\eta \equiv \rm{-ln\ tan}(\theta/2)$, and $\theta$ is the polar angle~\cite{coordinates}.  Surrounding
the silicon detector is a 3.1 m long gas and wire
drift chamber, the Central Outer Tracker (COT), which provides track position measurements in radial layers from 40 cm to 137 cm.  The COT has eight
superlayers, alternating between those parallel to the beam pipe and those at a $2^\circ$ stereo angle.  Each superlayer has 12 wires, providing up
to 96 track position measurements, for tracks with $|\eta| <$ 1. Good quality tracks are selected with cuts on the $\chi^2$ of the fit of the track
hits to a helical fit function.  Tracks traversing the full detector must have at least five position measurements in each of the four super-layers,
or three super-layers if the track does not traverse the whole detector. 

Surrounding the solenoid are sampling calorimeters with segmented towers projecting from the interaction region outward uniformly
in $\eta$. They measure the energy of interacting particles extending to a range of $|\eta| <$ 3.6. The inner electromagnetic calorimeters
have alternating lead and scintillator layers while the larger hadronic calorimeters are alternating iron and scintillator. Both types of calorimeters
sample the energy of the randomly fluctuating shower, leading to an uncertainty in the energy measurement. In electromagnetic cascades,
the primary energy loss mechanism is ionization, which provides a detectable signal in the scintillating layers.
Hadronic cascades however lose 30\% of their incident energy through the breakup of nuclei. Since this energy loss mechanism does not
yield a signal, energy resolution in the hadronic calorimeters is worse than in the electromagnetic calorimeter.  

\section{Standard Jet-Energy Corrections}
\label{jetcorrections}

Before reconstructed jets are used in a physics analysis, several jet-energy corrections for instrumental 
and showering effects are applied.  These are described in detail elsewhere~\cite{jetcorrs}, 
and are outlined below.

Currently CDF has nine levels of corrections, from level 0 through level 8. The standard set of corrections 
applied to Higgs analyses and also used in this analysis are the following : 

$\bullet$ {\bf Level 1} A relative correction is applied to the raw jet energies to make
the calorimeter response uniform in $\eta$. In general, the region $0.2<\mid\eta\mid<0.6$ is the
best understood. The response in the region is generally flat and the non-linearities are
well understood from test-beam measurements. The transverse energy of the two jets
in a 2$\rightarrow$2 process should be equal, and this property is used to scale the jet 
energies outside the $0.2<\mid\eta\mid<0.6$ region to the energy scale inside this region.

$\bullet$ {\bf Level 4} Multiple interaction correction. During the same bunch crossing,
the energy from different $p\bar p$-interactions can fall inside the jet cluster, increasing
the measured energy of the jet. The energy that needs to be subtracted is estimated
from the minimum bias data and is parameterized as a function of the number of vertices in the
event.

$\bullet$ {\bf Level 5} Absolute correction. After the relative corrections, the quark energies
are usually largely underestimated due to energy mis-measurement and nuclear absorption.
This correction factor is estimated from Monte Carlo simulations and test-beam data. However, it is
not enough to account for $b$ jets, whose energies are typically measured lower than other quark jets 
due to the higher probability of muons (and neutrinos) produced in the $b$ decay, which produce little 
(and no) signal in the calorimeter.

\section{Algorithms for $b$-Tagging}
\label{taggers}

\begin{figure}[H]
  \begin{center}
    \includegraphics[width=6.0cm,clip=]{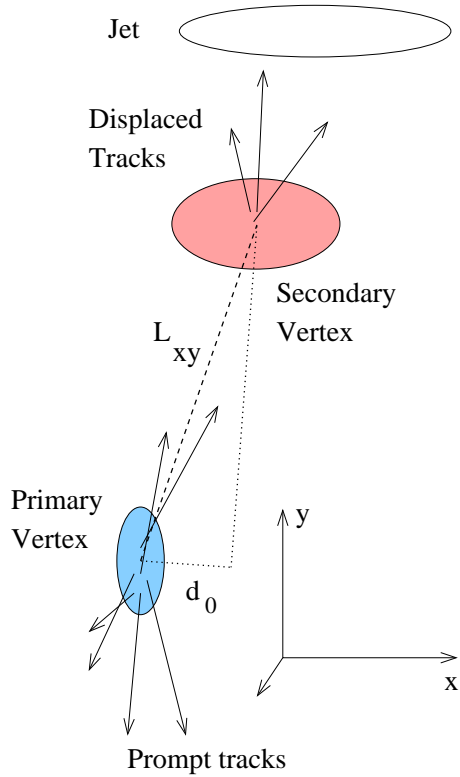}
    \caption{Figure demonstrating the reconstructed primary vertex, the secondary vertex
      and displaced tracks resulting from the $B$-hadron decay.}
    \label{btagging}
  \end{center}
\end{figure}

Quarks do not exist freely in nature and they undergo a process called hadronization in which they bound to other quarks. $B$ hadrons, although
unstable, have a relatively long lifetime of 1.5 picoseconds~\cite{pdg}, and are produced with typical transverse energies, (Et) of 50 GeV in Higgs
decays, providing them with relativistic boost factors of ($Et_b / m_b$) such that their lifetime in the lab frame can be ten times
larger, allowing them to travel several millimeters in a direction perpendicular to the beam. When the $B$-hadrons decay, their mass propels
the tracks in slightly different directions relative to the direction of the initial $B$ hadron, allowing a secondary vertex to
be reconstructed using tracks a few millimeters away from the beam~\cite{btagging}. These tracks do not intercept the
beam when extrapolated
backward and have high impact parameters as shown in Figure \ref{btagging}. The CDF experiment provides excellent tracking with a precision silicon
vertex detector~\cite{svx}, and several algorithms using this detector have been developed to distinguish $B$-hadron decays
from light-hadron decays.

The secondary-vertex tagging algorithm identifies $B$ hadrons by calculating a secondary vertex position from tracks originating
near the primary vertex. This method has been used in other Higgs boson searches and in most of the top quark measurements.
When the significance of $\sigma_{L_{xy}}/L_{xy}$ of the decay lenght, $L_{xy}$, in transverse plane is 6.5 sigma, we identify the
jet as secondary-vertex tagged (ST). In the $WH$ analysis we expect
two $b$ jets originating from the Higgs decay. Other $b$-tagging algorithms are used in the $WH$ analysis to maximize efficiency
for detecting $b$-jet decays, but most of the sensitivity comes from the ST-tagger, and in this paper we introduce a method to
correct the ST-tagged $b$-jet energies.

\section{Measurements Correlated to $b$-jet Energy Measurement}

To produce a correction that will improve both the $b$-jet energy measurement and its resolution, 
we consider variables that are correlated with the $b$-quark energy and are well measured.
Variables that are correlated to the quark energy fall into three classes: energy
measurements from the calorimeter, track momentum measurements from the COT, and vertex displacement measurements from extrapolating
tracks into the SVX+ISL. Calorimeter-energy measurements are the standard way of measuring jets as discussed above. Jets are mostly composed
of hadrons such as $\pi^+$ and $\pi^-$, which deposit energy mainly in the hadronic calorimeter, which has a
resolution of $\sigma(E)/E \sim 50\%/\sqrt{E}$.
On average $\frac{1}{3}$ of the jet energy is carried by $\pi^0$ particles, which decay to two photons, leaving most of the energy in the
electromagnetic calorimeter, which has a resolution of $\sigma(Et)/Et = 13.5\%/\sqrt{Et} \oplus 2\%$. Tracking momentum measurements are
highly correlated to quark energy, since on average $\frac{2}{3}$ of the particles in jets are charged particles, and the
tracker has a high efficiency
and a precise momentum resolution of $\sigma(Pt)/{Pt} \sim 0.15\% \cdot Pt$.
Assuming the $B$ hadron is moving at the speed of light, the lifetime $\tau$ is proportional to the
decay length (L) in the lab frame, and the energy and the transverse energy are determined by 

\begin{equation} E  =  L \cdot (m / c\tau) \rm{,} \end{equation}

and

\begin{equation} Et  =  L_{xy} \cdot (m / c\tau) \rm{,}  \end{equation}

where $m$ is the mass of the $B$ hadron.

The secondary vertex resolution is about 40 $\mu$m and the $B$-hadron lifetime is about
470.1$\pm$2.7$\mu$m~\cite{pdg}.
Before $b$-tagging, for low energy $B_u$ decays, $\sigma_{L_{xy}}/L_{xy}$ is measured in data to be about 44\%. 
For a 50 GeV $B$ hadron, it will travel almost a half centimeter before decaying and producing a secondary 
vertex.  After $b$-tagging jets by requiring $L_{xy} / \sigma_{L_{xy}} >$ 6.5, the remaining events have a  
$\sigma_{L_{xy}}/L_{xy}$ of about 4\%.

\section{Variables Used in the $b$-jet Correction Function}
\label{inputs}

We studied a pool of variables calculated for jets matched to $b$ quarks in fully-simulated
Monte Carlo events and validated the modelling by comparing to jets selected 
in data events that were $b$-tagged by the above algorithm. From the variables that 
were well-modeled in the data, we determined a minimal set of variables that 
could improve jet-energy resolution. We use these variables to develop a correction function to determine the true $b$-quark energy. 

For each jet, we studied 40 variables related to 
the calorimeter energies, the charged tracks, and the displaced vertices of the jets of
cone $\Delta R = \sqrt{\Delta\eta^2 + \Delta\phi^2} < 0.4$
and converged on nine variables most optimal for the jet-energy correction. The four calorimeter variables chosen are the jet $Et$ before
corrections (raw jet $Et$), the jet $Et$ and $Pt$ after the standard jet corrections (Level 5 jet $Et$, Level 5 jet $Pt$)
and the tranverse mass
of the jet, defined as $(Et/E)\cdot M$, where M is a jet mass from the jet clustering algorithm.  The corrected
jet $Et$ provides the single best estimate of the true quark energy. The raw jet $Et$ is important because its difference with
the corrected jet $Et$ indicates how well the jet was measured: jets which are highly corrected are likely in regions of the detector
that are less instrumented. The degree to which $Pt$ of the jet is different than the $Et$ of the jet indicates how collimated the
jet is. A smaller difference corresponds to a more collimated jet, which is an indication how well contained the jet is by the
fixed cone algorithm. The transverse mass of the jet also provides
similar information.  The tracking variables chosen are the sum of the $Pt$ and the maximum $Pt$ track of the set of
good quality tracks in the jet cone.
Sum $Pt$ is an excellent measurement of the energy of charged particles within the jet, which carry about $\frac{2}{3}$ of the jet energy.
The maximum $Pt$ of good quality tracks, as compared to the sum, provides information about how well the track energy is distributed among
the particles and the likelihood of overlapping particles in the jet.  These tracks are not required to have silicon hits attached.
Tracks with much higher $Pt$ than the calorimeter energy are excluded as they likely result from a mismeasurement of track parameters.
Tracks with $Pt <$ 1 GeV are not included in order to improve the modeling and reduce the effect of pile up of additional interactions in the same bunch
crossing. The vertex variables used are the reconstructed secondary-vertex distance in the transverse plane ($L_{xy}$) with respect to
the primary vertex position, the uncertainty on $L_{xy}$, ($\sigma_{L_{xy}}$) and the fitted secondary vertex Pt (SecvPt).
These variables, as discussed above, provide an independent
estimate of the value and its uncertainty of the $b$-quark $Et$. For the secondary vertex variables we require silicon hits in order to
determine better the vertex position, whereas for the tracking variables, we do not require silicon hits to improve the reconstruction efficiency.  

\begin{figure}[H]
  \begin{center}
    \includegraphics[width=6.0cm,clip=]{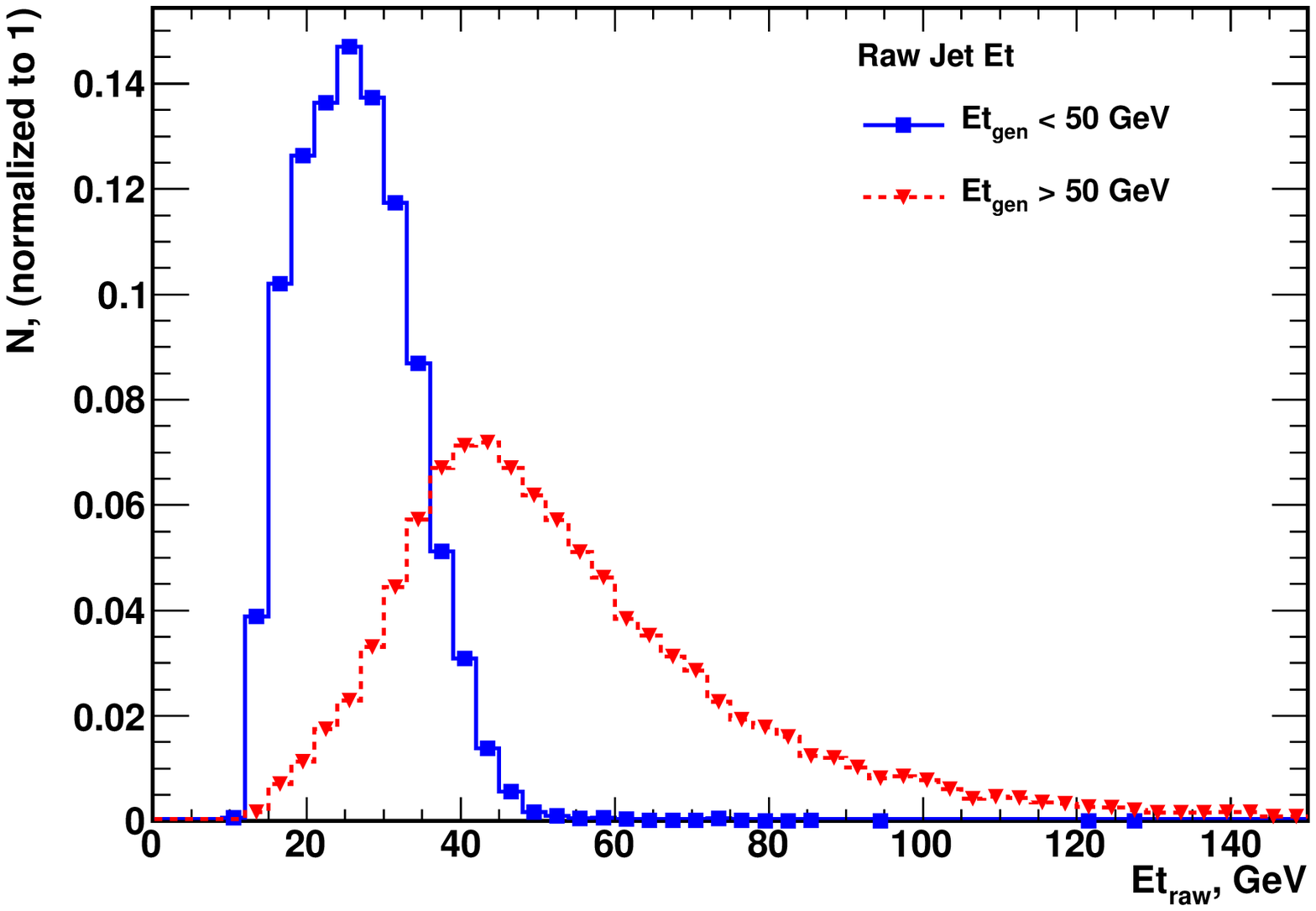}
    \includegraphics[width=6.0cm,clip=]{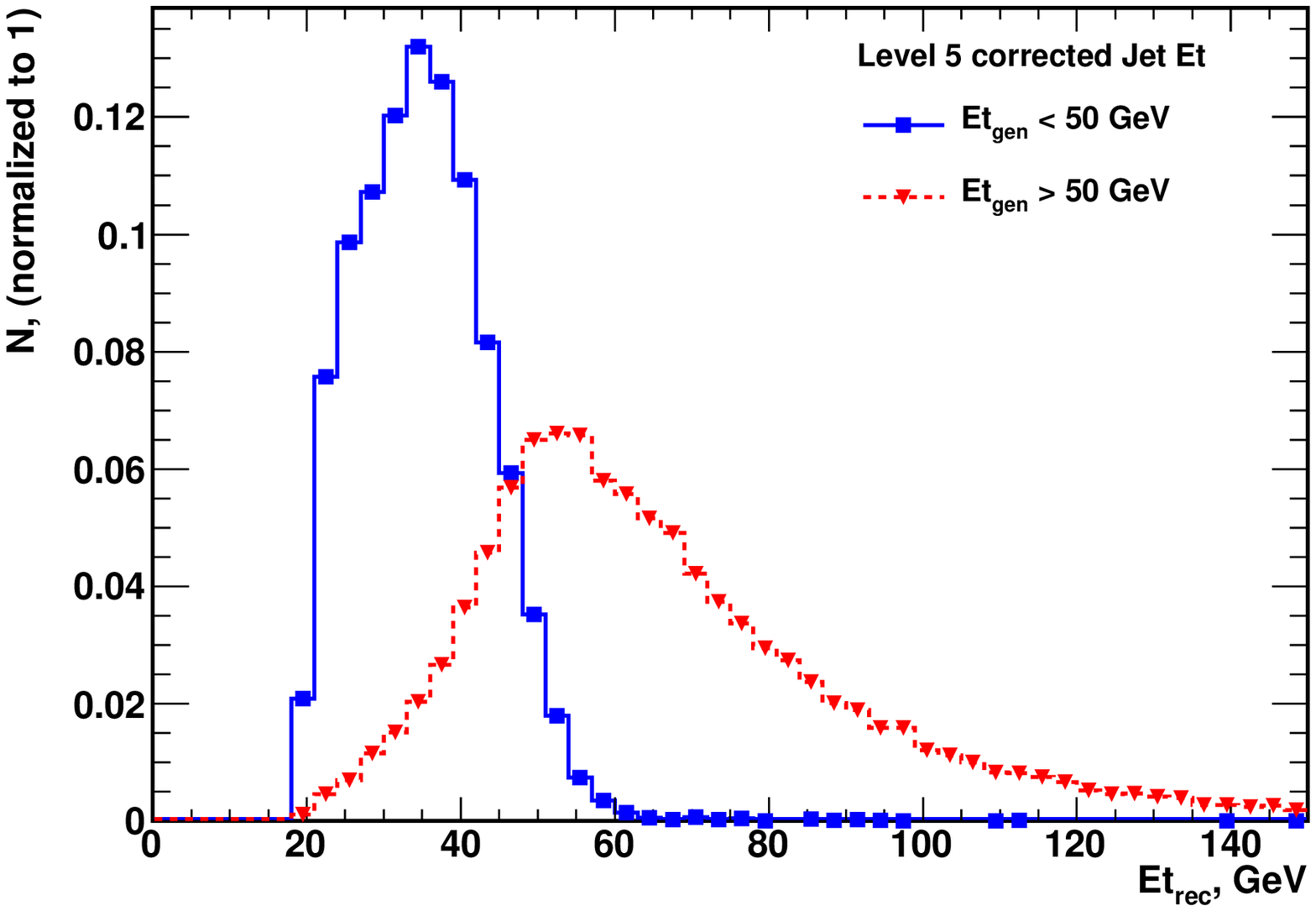}
    \caption{The left plot shows that the measured raw jet Et is correlated with the generated
      true quark energy. The blue (solid) line shows the raw jet Et for generated quarks Et $<$ 50 GeV,
      the red (dotted) line for Et $>$ 50 GeV. The right plots shows the correlation for the level 5
      corrected jet Et.}
    \label{gcorr1}
  \end{center}
\end{figure}

\begin{figure}[H]
  \begin{center}
    \includegraphics[width=6.0cm,clip=]{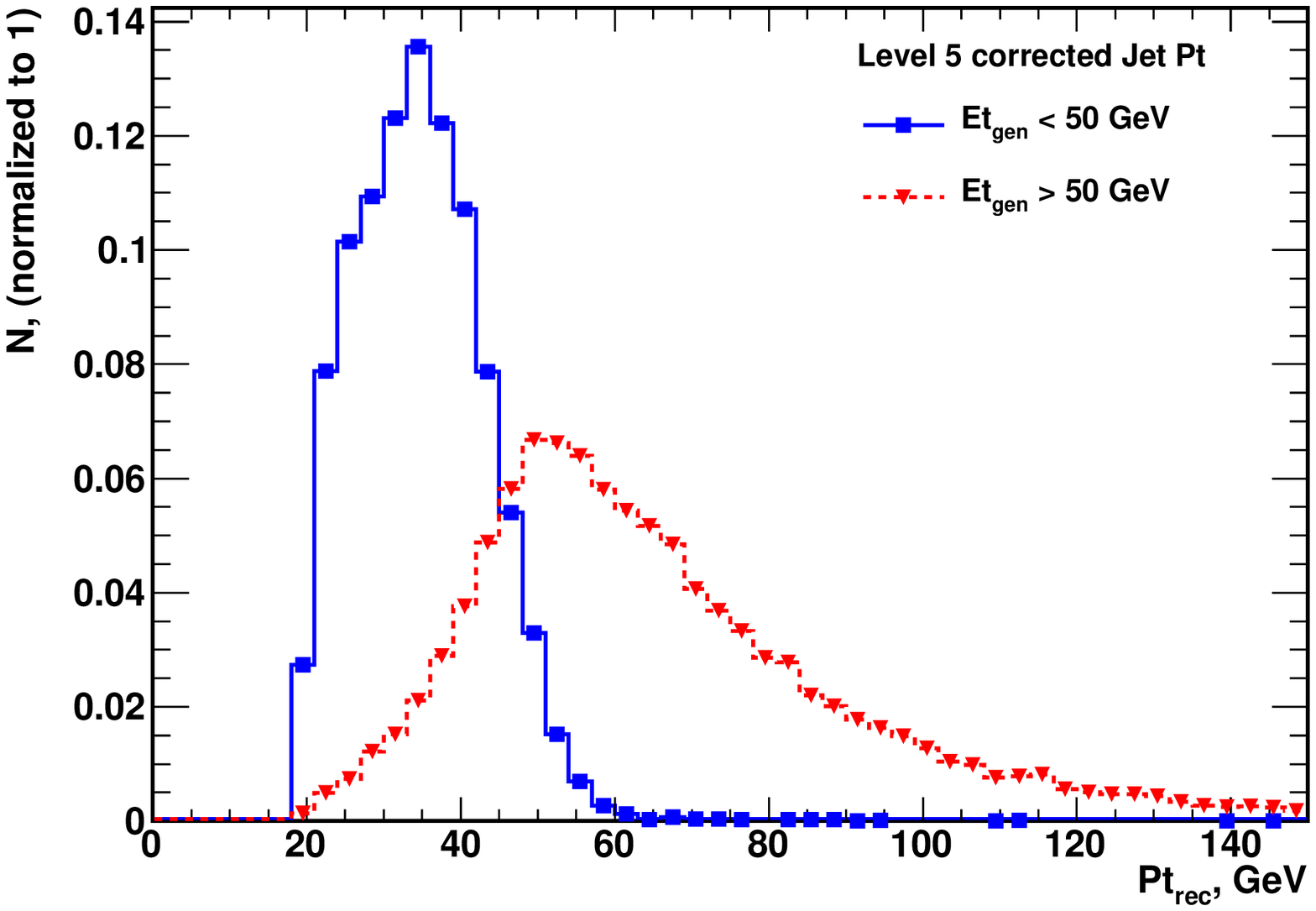}
    \includegraphics[width=6.0cm,clip=]{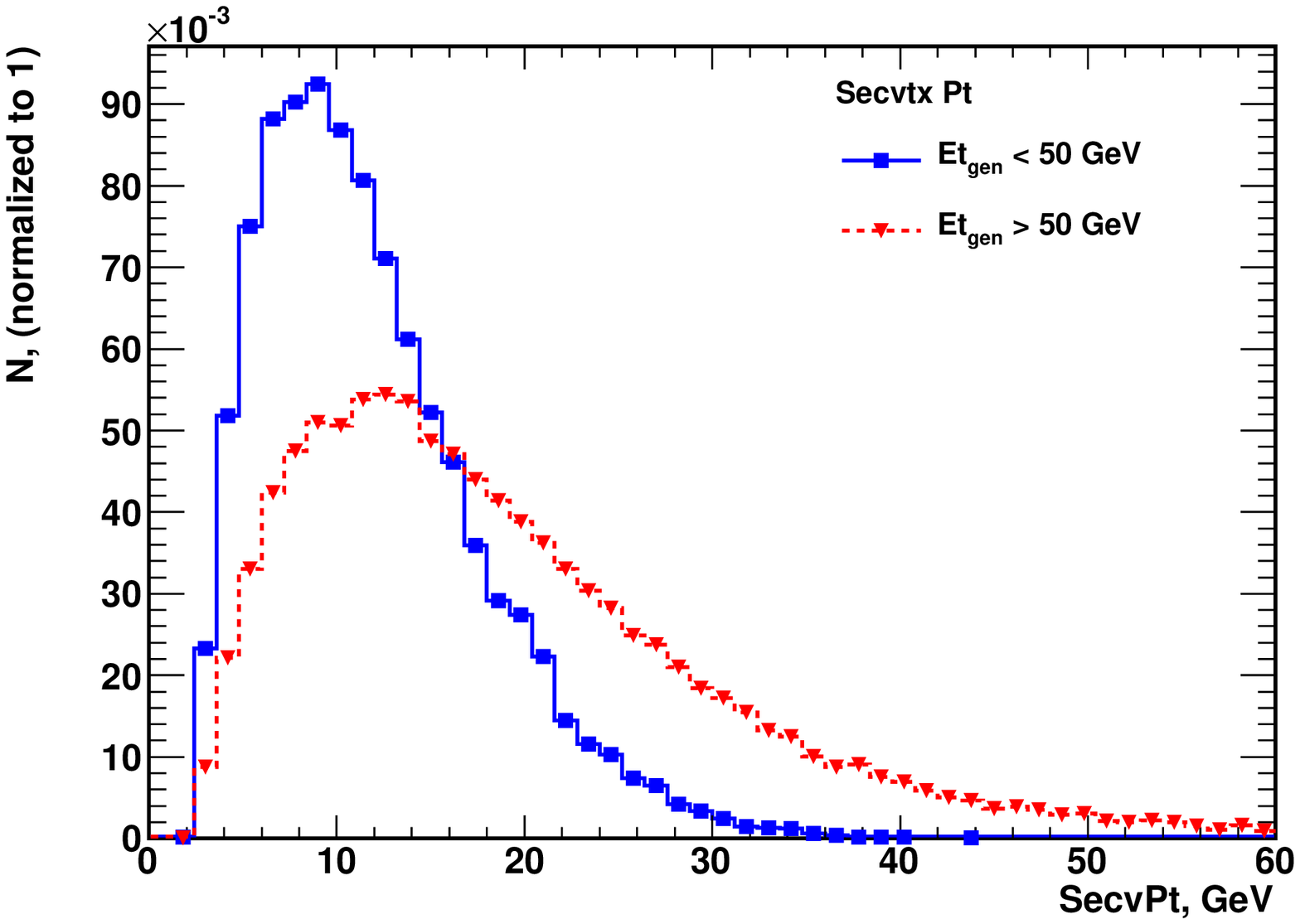}
    \caption{The left plot shows that the level-5 corrected jet Pt is correlated with the generated
      true quark energy. The blue (solid) line shows the level-5 corrected jet Pt for generated quarks Et $<$ 50 GeV,
      the red (dotted) line for Et $>$ 50 GeV. The right plots shows the correlation for the fitted secondary
      vertex Pt.}
      \label{gcorr2}
  \end{center}
\end{figure}

\begin{figure}[H]
  \begin{center}
    \includegraphics[width=6.0cm,clip=]{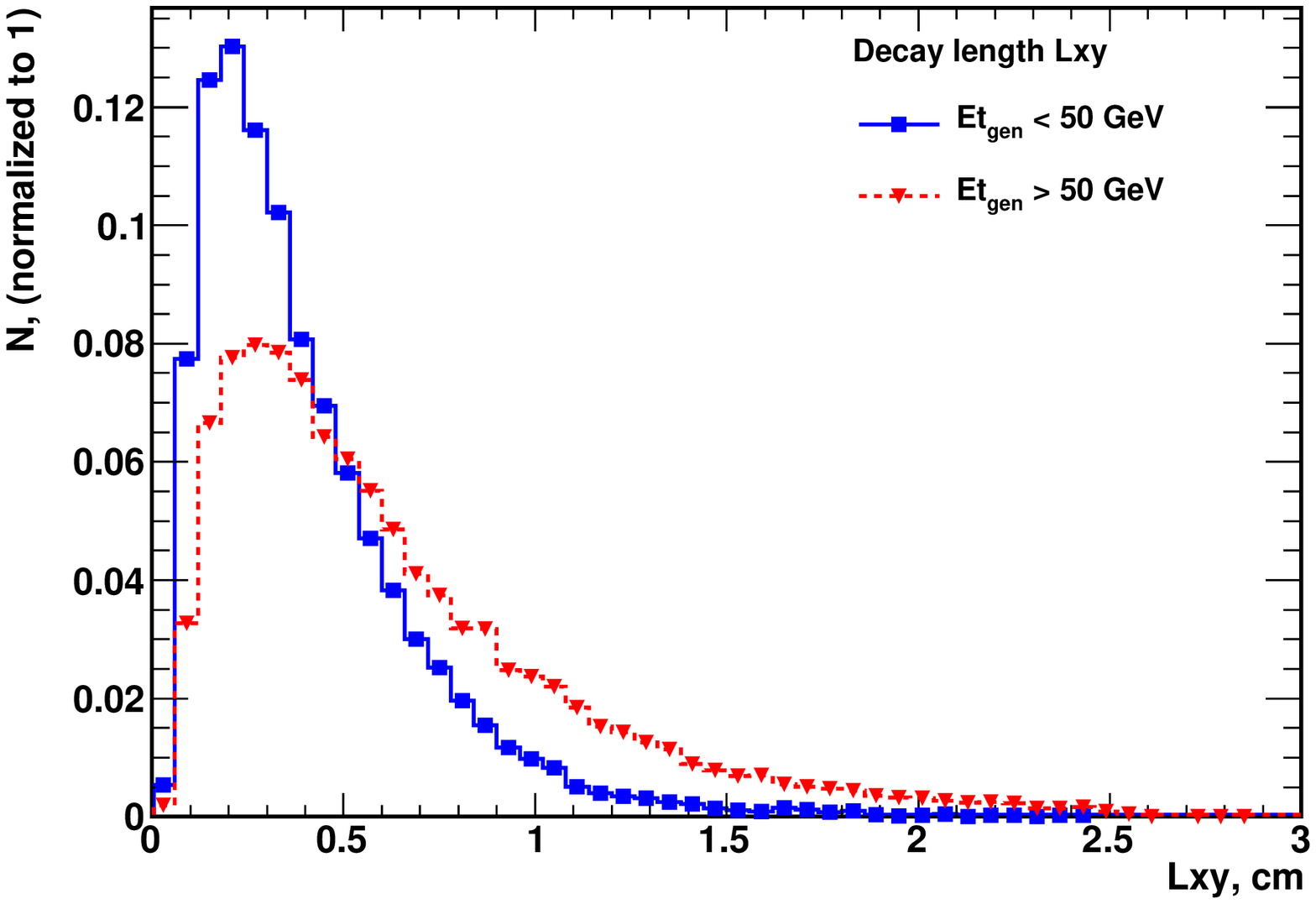}
    \includegraphics[width=6.0cm,clip=]{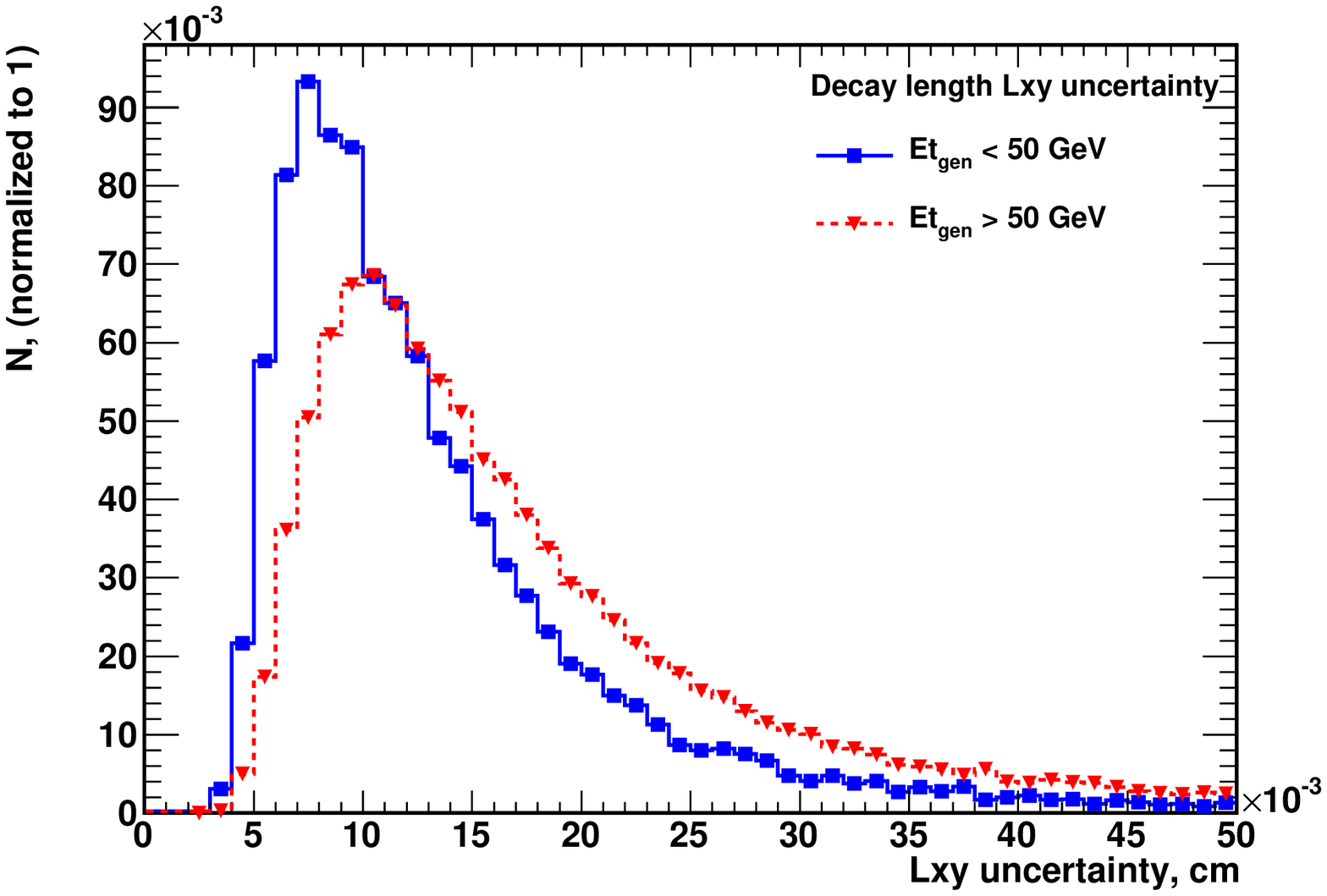}
    \caption{The left plot shows that the measured secondary vertex position in the xy-plane has a
      correlation with respect to the generated true quark enertgy.  The blue (solid) line shows the secondary
      vertex position in the xy-plane for generated quarks Et $<$ 50 GeV,
      the red (dotted) line for Et $>$ 50 GeV. The right plots shows the correlation for the
      uncertainty of the secondary vertex position.}
      \label{gcorr3}
  \end{center}
\end{figure}

\begin{figure}[H]
  \begin{center}
    \includegraphics[width=6.0cm,clip=]{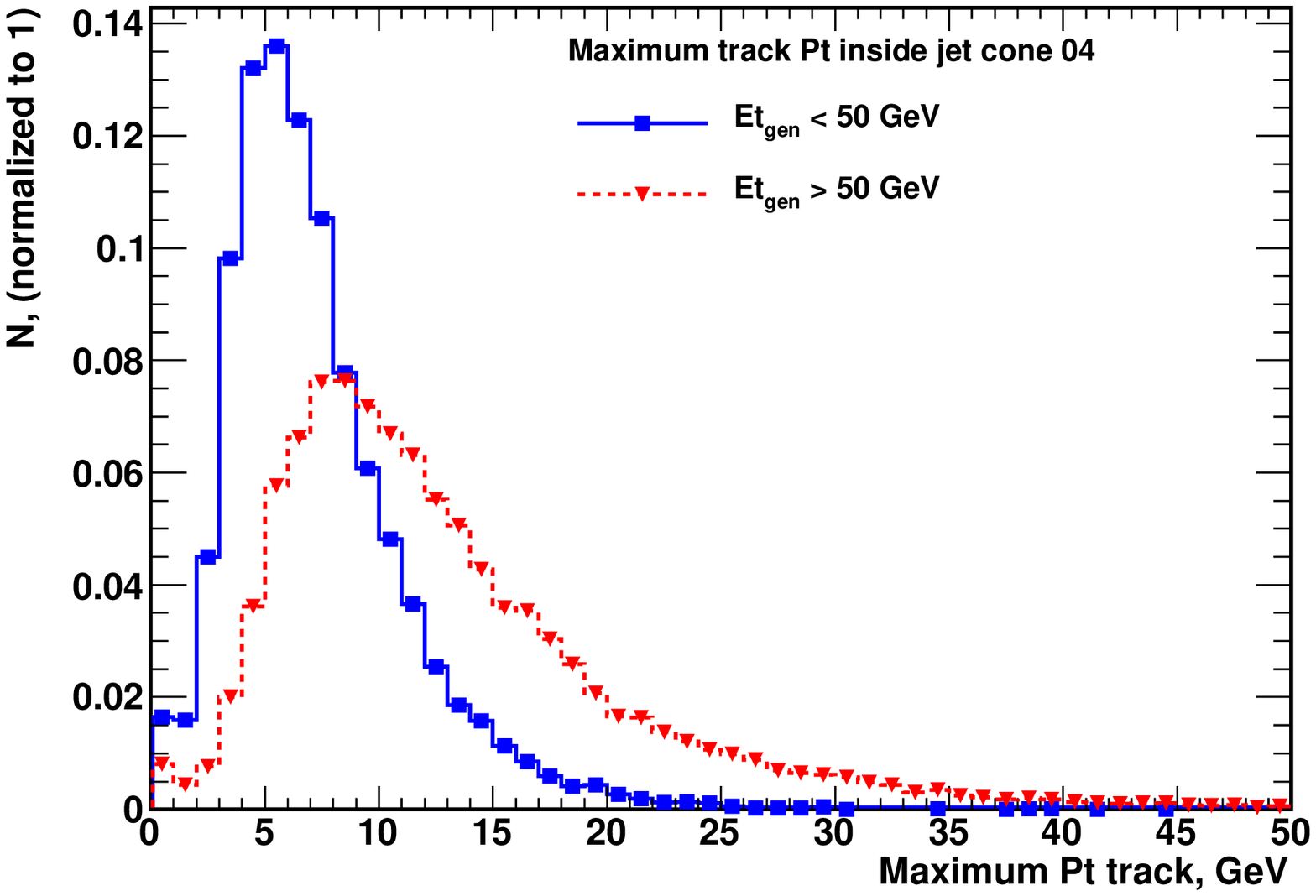}
    \includegraphics[width=6.0cm,clip=]{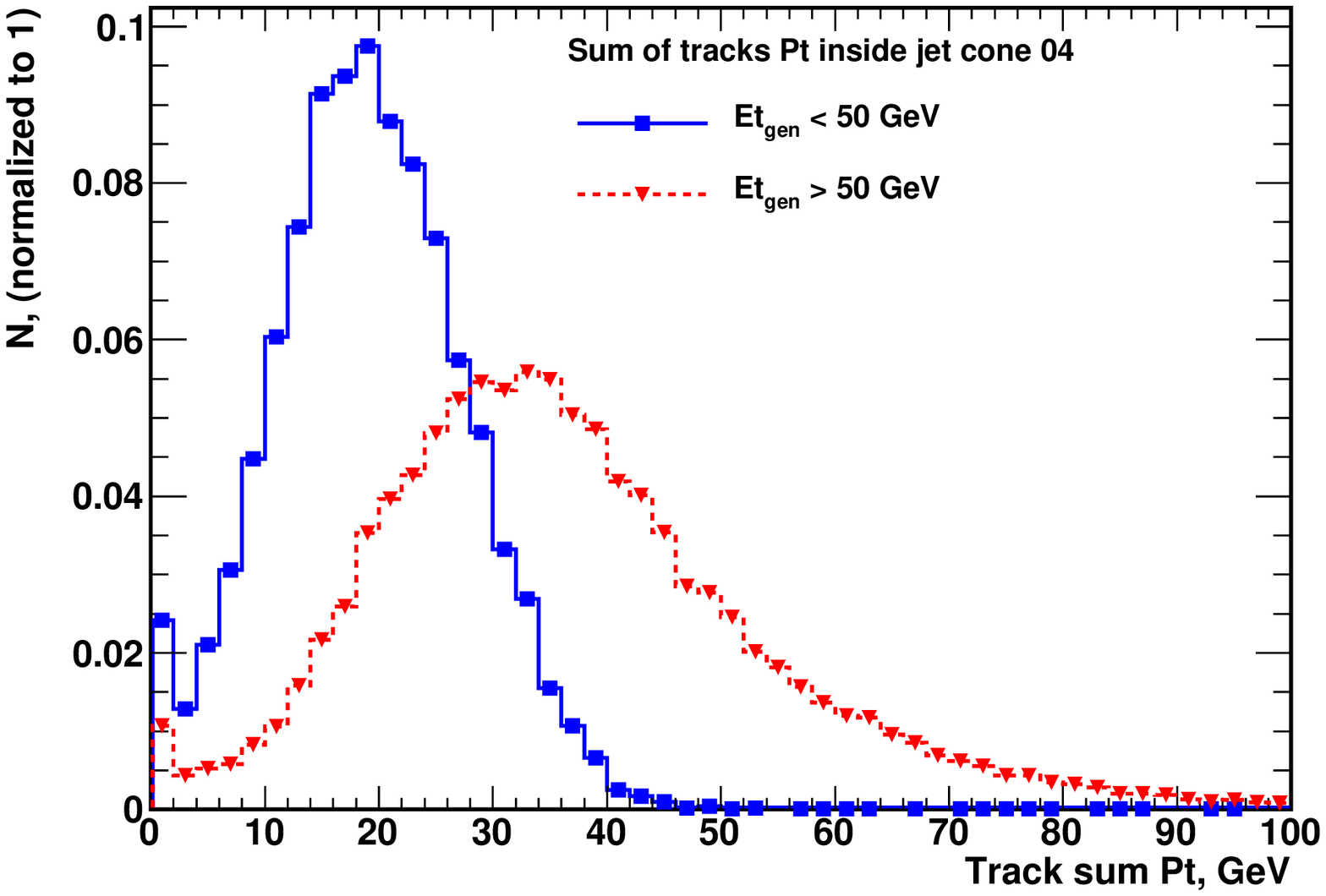}
    \caption{The left plot shows that the maximum Pt of the good quality track inside the jet cone 04 has a
      correlation with respect to the generated true quark enertgy.  The blue (solid) line shows the
      maximum Pt of the good quality track inside the jet cone  for generated quarks Et $<$ 50 GeV,
      the red (dotted) line for Et $>$ 50 GeV. The right plots shows  the correlation for the
      sum of Pt of the all good quality tracks inside jet cone.}
      \label{gcorr4}
  \end{center}
\end{figure}

\begin{figure}[H]
  \begin{center}
    \includegraphics[width=6.0cm,clip=]{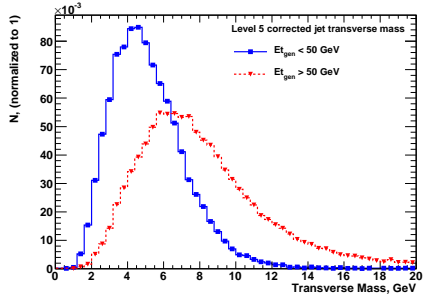}
    \caption{The plot shows that the jet transverse mass is correlated with the generated
      true quark energy, the blue (solid) line shows the jet transverse mass for generated quarks Et $<$ 50 GeV,
      the red (dotted) line for Et $>$ 50 GeV.}
      \label{gcorr5}
  \end{center}
\end{figure}

Figures \ref{gcorr1}-\ref{gcorr5} show each of the nine variables for a Monte-Carlo-generated Higgs mass sample of 115 GeV/$c^2$ for
generated $b$-quark energy $Et_{gen} < $ 50 GeV and  $Et_{gen} > $ 50 GeV, after $b$-tagging with the $ST$-algorithm.
The difference between the shapes for the two energy regions indicates how correlated the variable is with the true quark energy. 


\section{Monte Carlo Validation}
\label{validation}

\begin{figure}[H]
  \begin{center}
    \includegraphics[width=6.0cm,clip=]{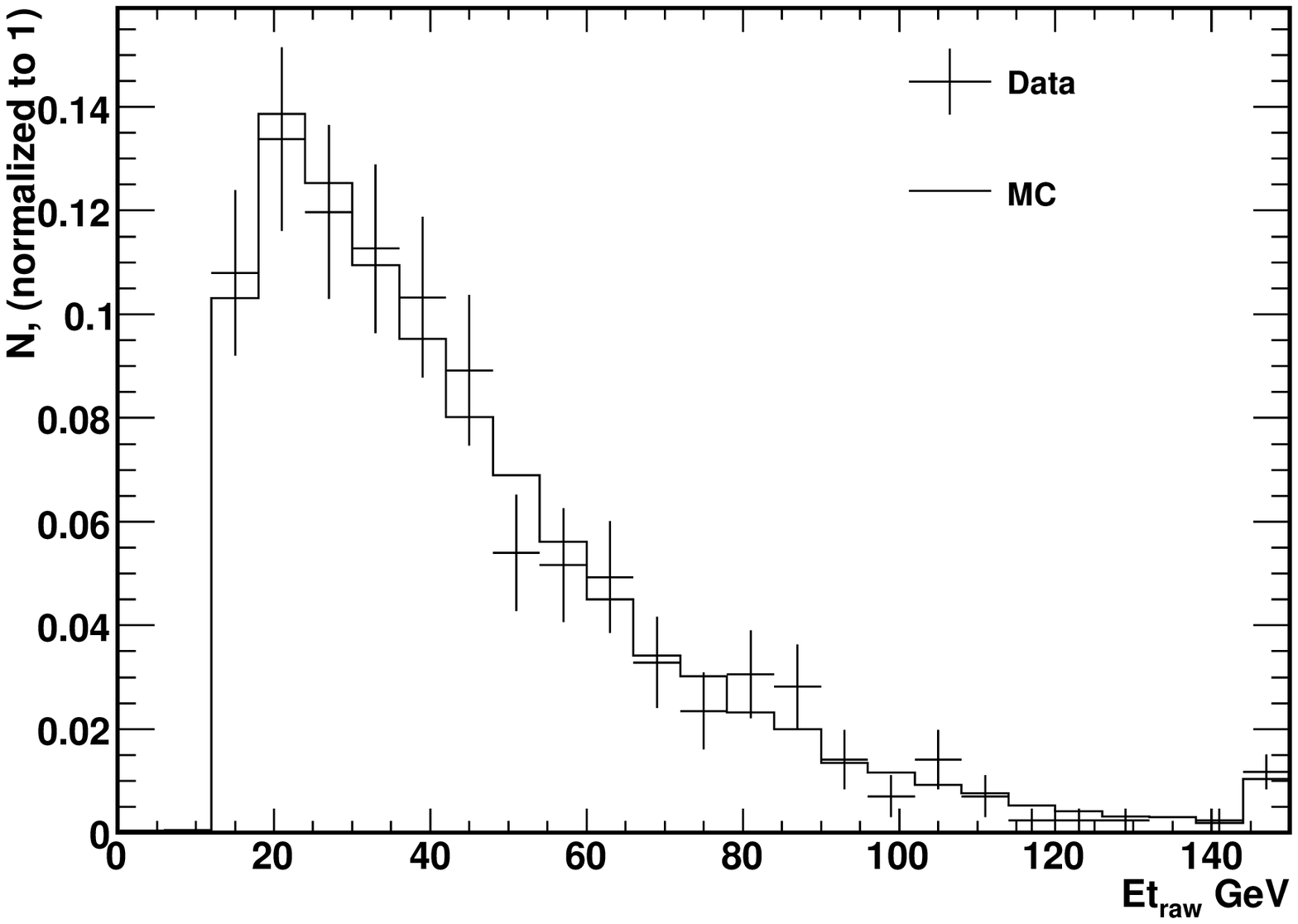}
    \includegraphics[width=6.0cm,clip=]{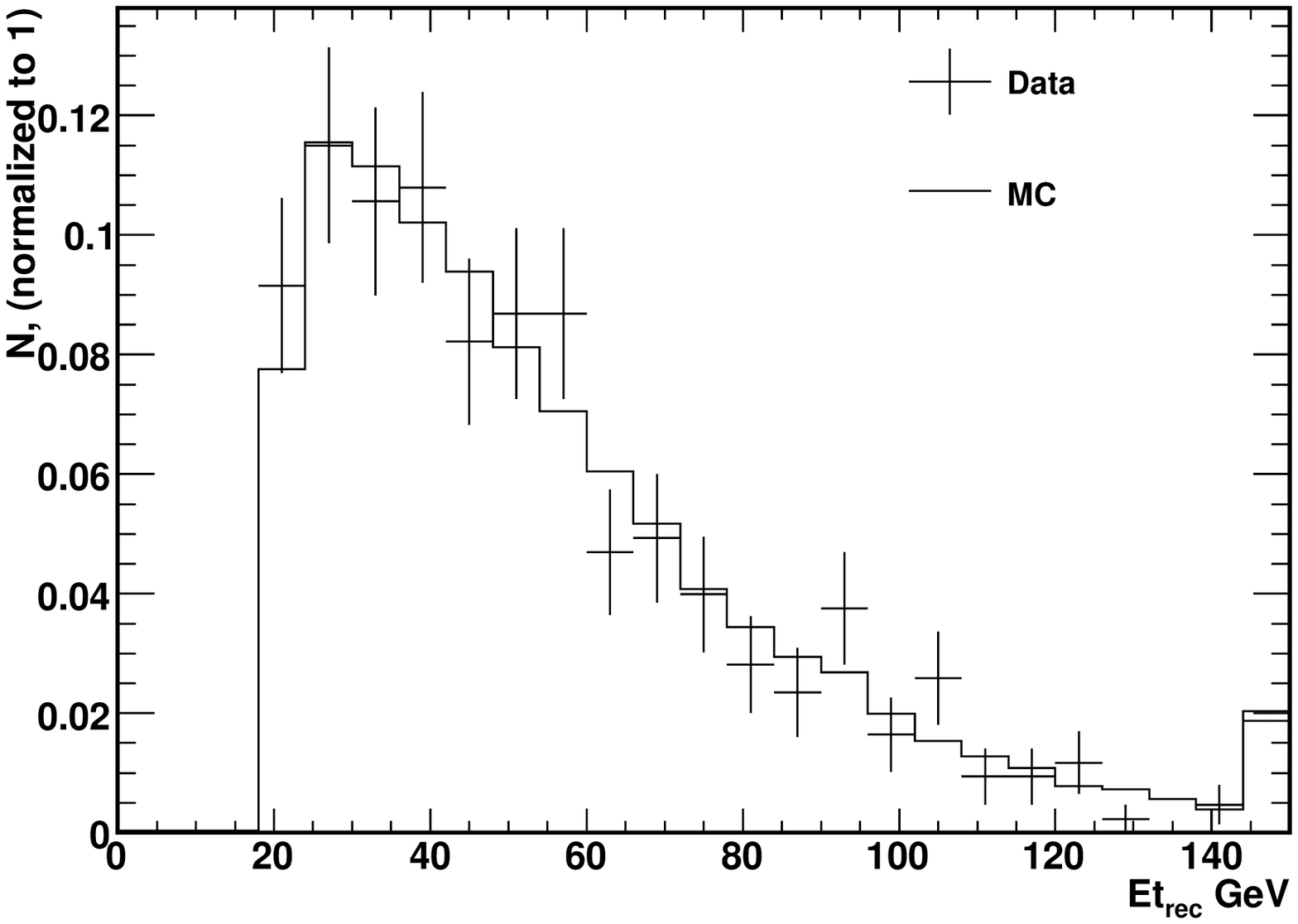}
    \caption{The left plot shows the data MC comparison of the uncorrected raw jet Et, the right plot
      shows the comparison for the level-5 corrected jet Et.}
    \label{val1}
  \end{center}
\end{figure}

\begin{figure}[H]
  \begin{center}
    \includegraphics[width=6.0cm,clip=]{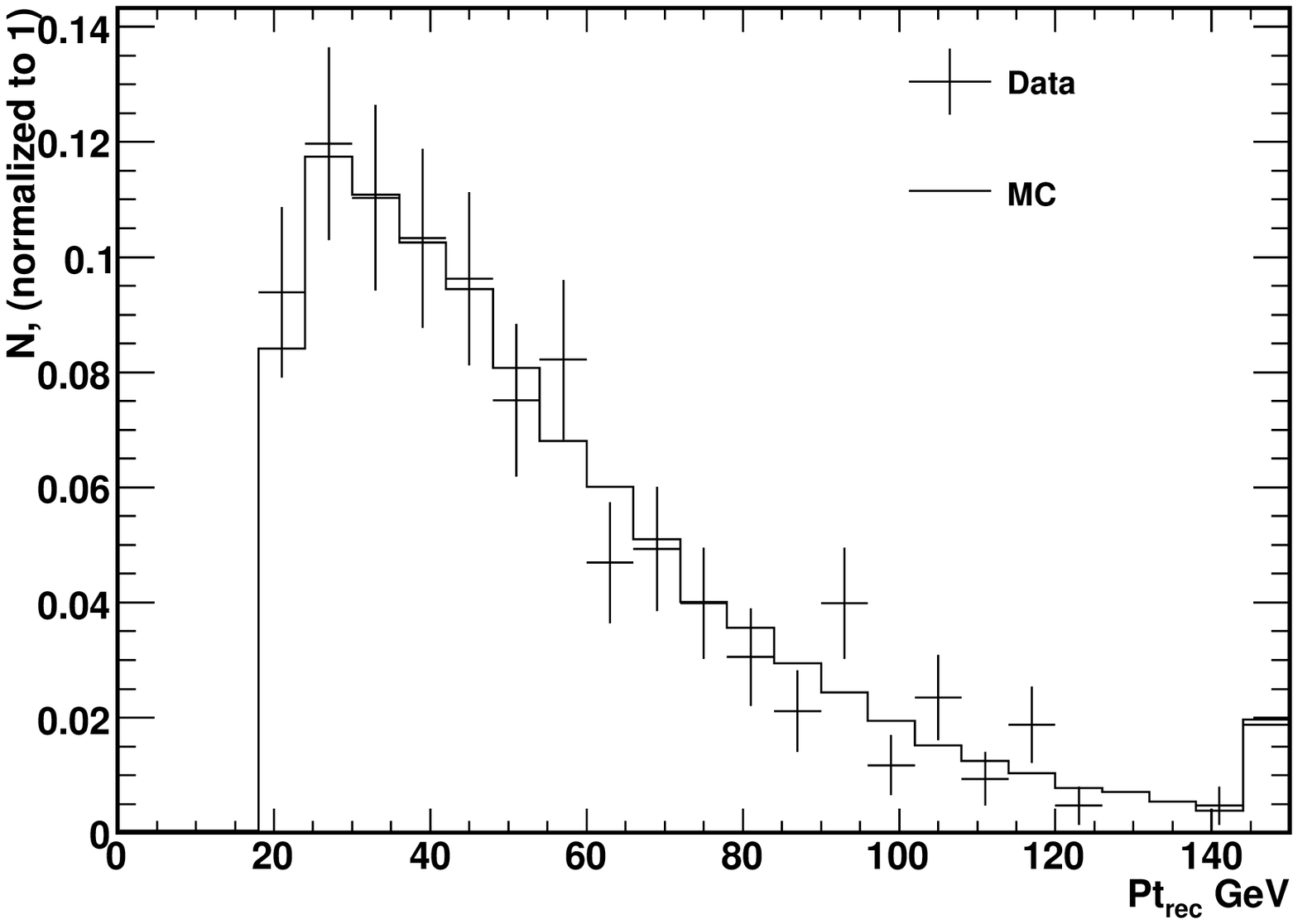}
    \includegraphics[width=6.0cm,clip=]{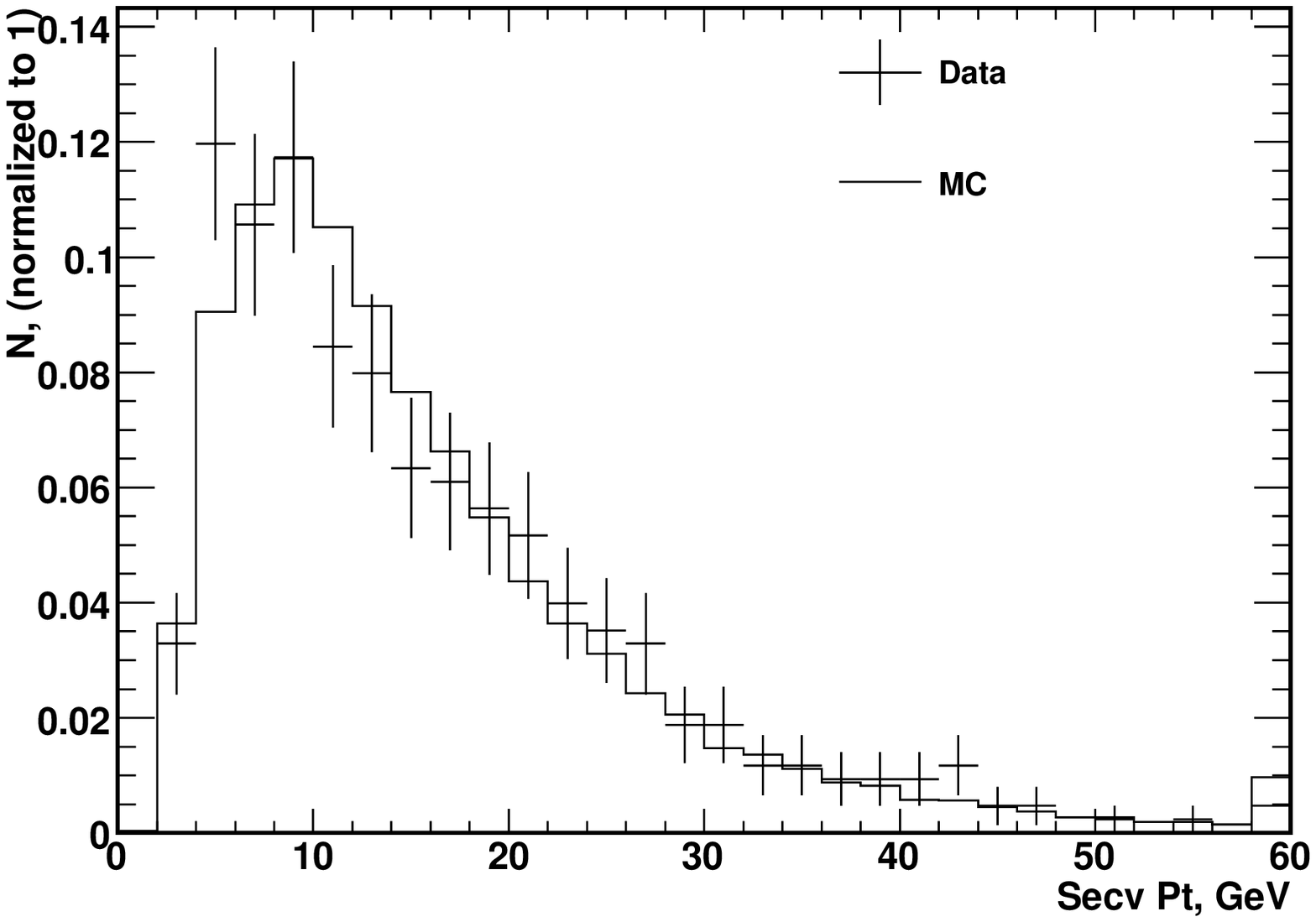}
    \caption{The left plots shows the data MC comparison of the level-5 corrected jet Pt, the right plot
      shows the comparison for the fitted secondary vertex Pt.}
    \label{val2}
  \end{center}
\end{figure}

\begin{figure}[H]
  \begin{center}
    \includegraphics[width=6.0cm,clip=]{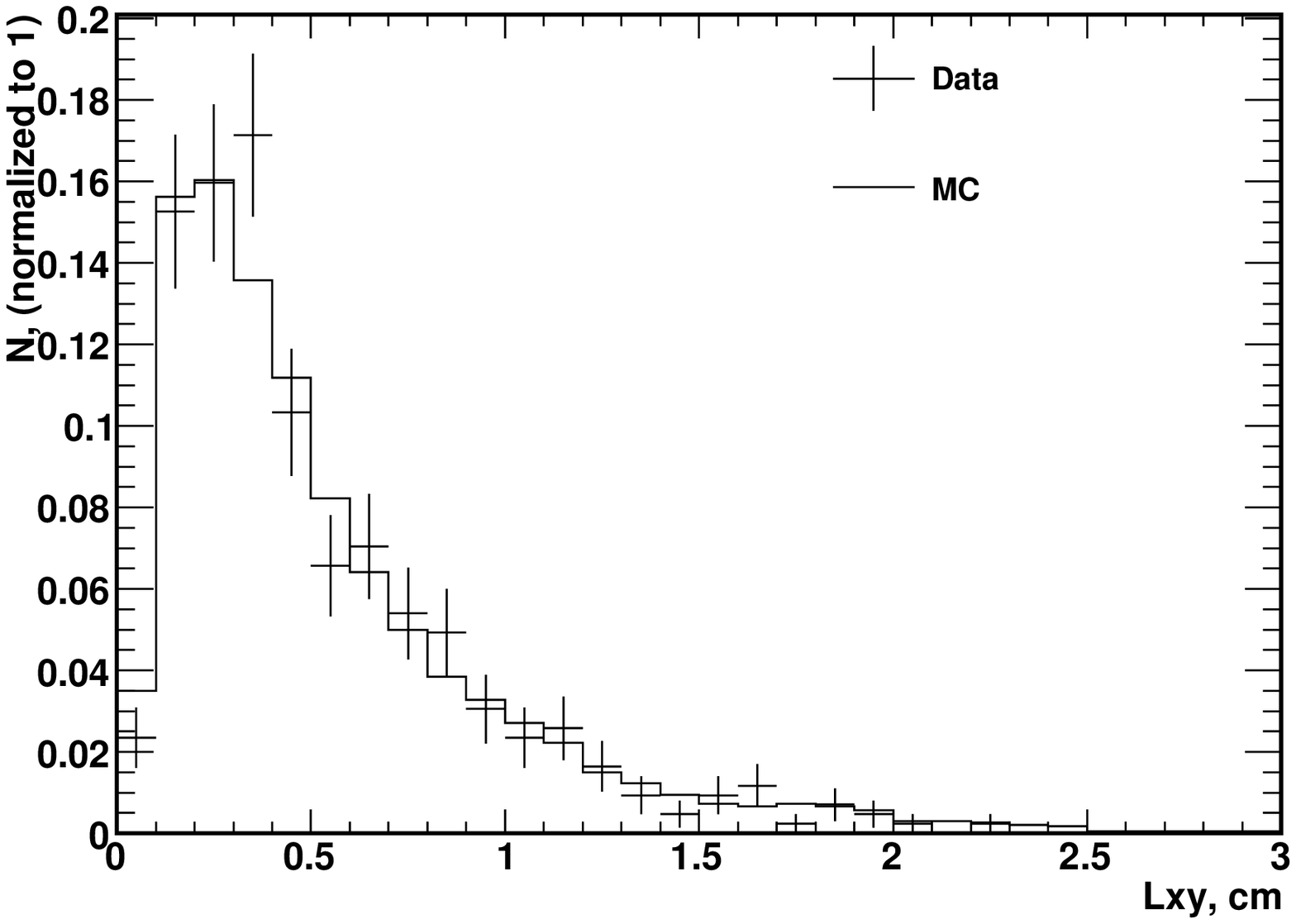}
    \includegraphics[width=6.0cm,clip=]{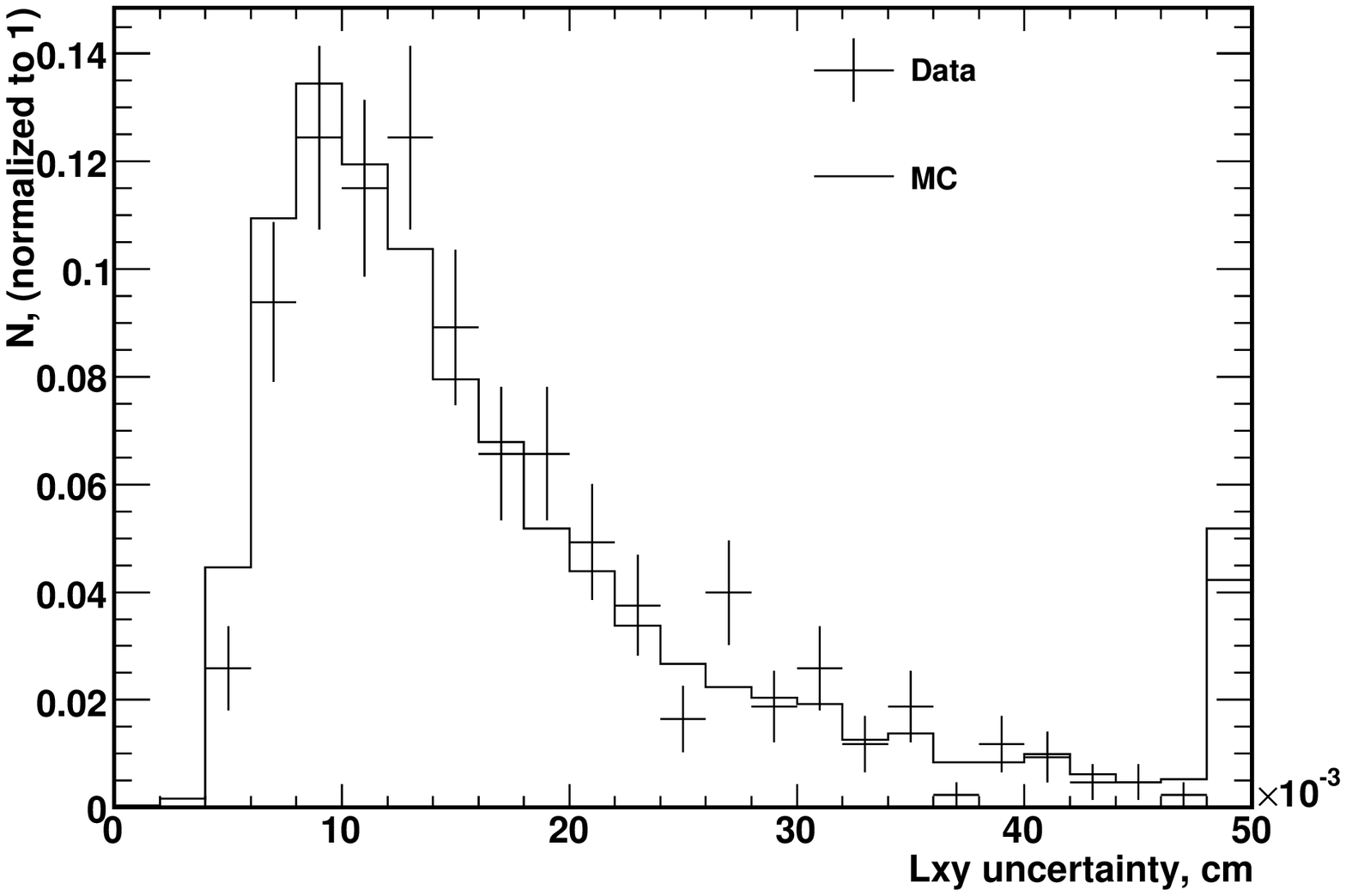}
    \caption{The left plots shows the data MC comparison of the measured secondary vertex position in the xy-plane,
      the right plot shows the comparison for the vertex position uncertainty.}
    \label{val3}
  \end{center}
\end{figure}

\begin{figure}[H]
  \begin{center}
    \includegraphics[width=6.0cm,clip=]{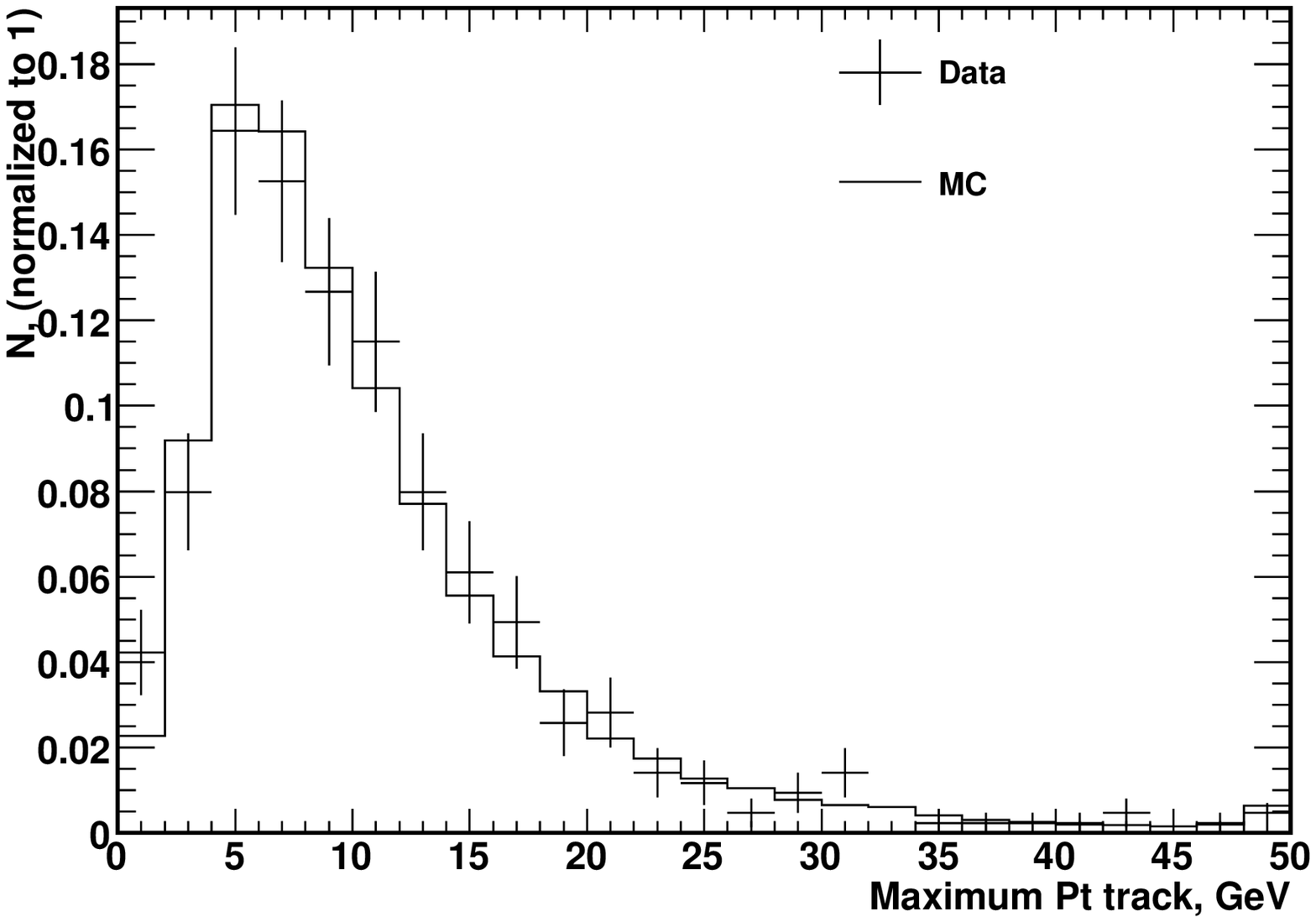}
    \includegraphics[width=6.0cm,clip=]{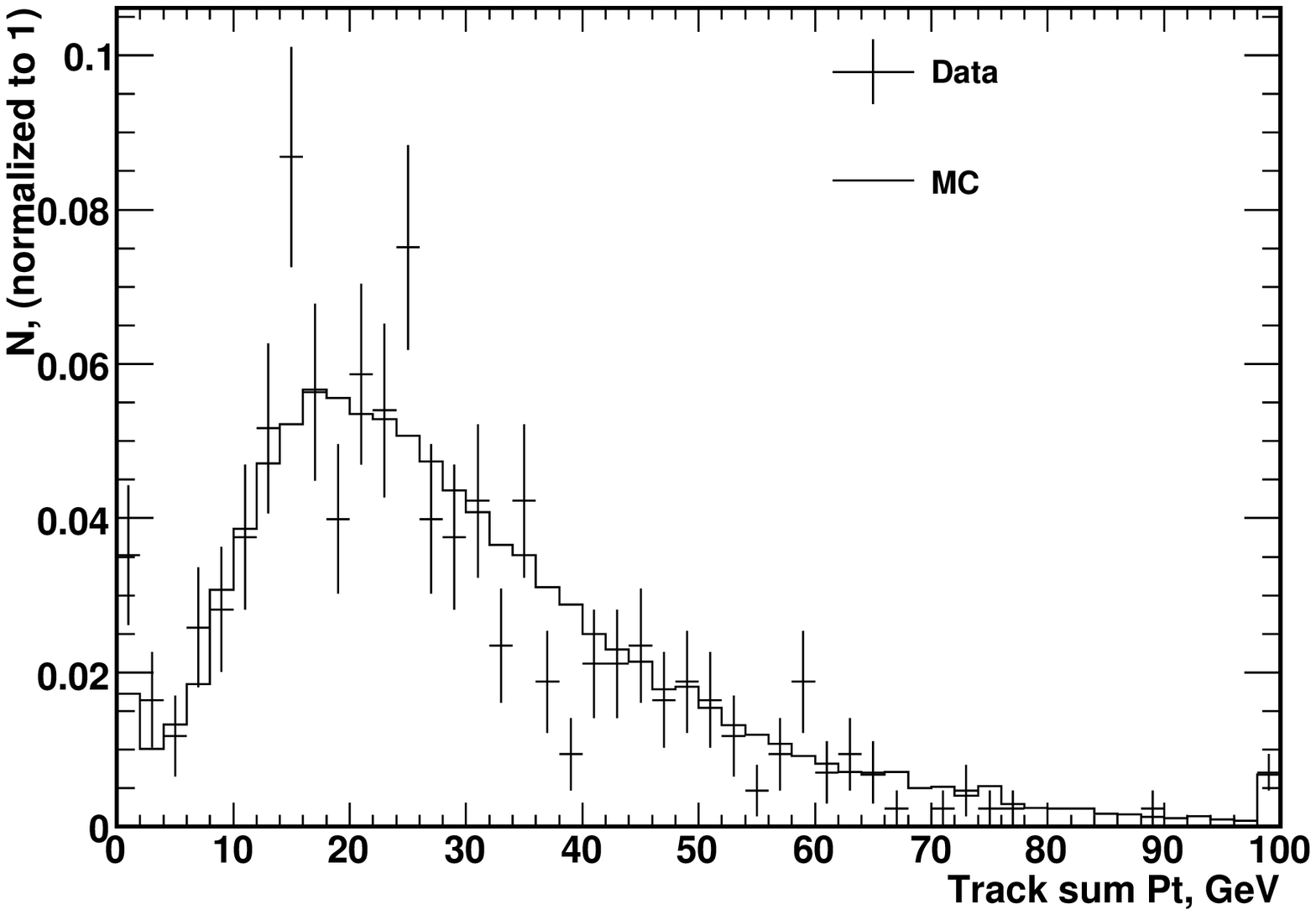}
    \caption{The left plots shows the data MC comparison of the maximum $Pt$ track inside a jet, the right plot
	shows the comparison for the sum of the track $Pt$ inside a jet.}
    \label{val4}
  \end{center}
\end{figure}

\begin{figure}[H]
  \begin{center}
    \includegraphics[width=6.0cm,clip=]{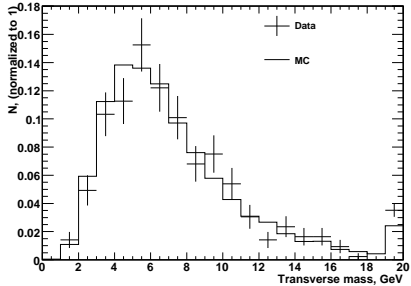}
    \caption{The plot shows the data MC comparison of the transverse jet mass.}
    \label{val5}
  \end{center}
\end{figure}

Validation of the modeling in data of the above variables is done in events where two jets are both tagged by the ST $b$-identification
algorithm. Jets are sorted with respect to their energy transverse to the beam $Et$ such that the jet with higher $Et$ is
called the leading jet, and the next highest is the sub-leading jet. We combine the leading and the sub-leading jet samples to double
the available statistics to 426 jets. Validation plots
are shown on Figures \ref{val1}-\ref{val5} and present a good agreement between data and Monte Carlo. For higher statistics
validations in a $b$-enriched sample, we refer to studies that were done in the context of the CDF single-top discovery, which used many
variables including the ones used in this paper, in order to remove light quark and charm jets from the background in
order to enhance the single-top signal ~\cite{kit}.
 
\section{Correction Function for $b$-jet}
\label{training}

\begin{figure}[H]
  \begin{center}
    \includegraphics[width=6.0cm,clip=]{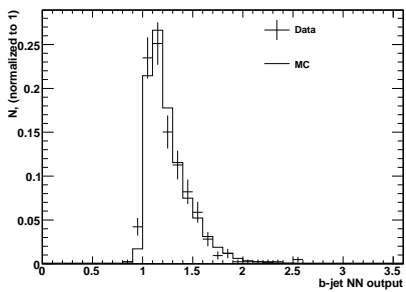}
    \caption{The plot shows the correction value determined by the Neural 
      Network, which is applied as a scale factor to the level-5 corrected $b$-jets four vector.}
    \label{nnout}
  \end{center}
\end{figure}

Using the nine discriminating variables discussed above, we develope a multivariate 
function to calculate the true $b$-quark energy, in order to produce an improved jet-energy 
resolution.  We train an Artificial Neural Network (NN) to estimate correctly the quark energy in a sample of $b$-tagged jets
chosen from a Monte Carlo sample of $WH$ events with the Higgs boson forced to decay to a $b$ quark and anti-$b$ quark.
In order to avoid a bias of the jet-energy correction with respect to the Higgs boson mass in 
the training, we train the NN on a sample of $WH$ Monte-Carlo-simulated events containing equal amount of events generated with masses equally
sampled in 10 GeV/$c^2$ increments from 100 - 150 GeV/$c^2$.
Since this is the range of the CDF $WH$ analysis, this insures that the correction function does not learn to correct jets to obtain a particular Higgs
boson mass.  The correction function utilized a Multi-Layer Perceptron (MLP) Neural Network implemented in
ROOT~\cite{ROOT}. The training is done with ROOT's implementation of Broyden, Fletcher, Goldfarb, Shanno (BFGS) method with 500 epochs.
We use nine input variables, as summarized in the section \ref{inputs}, with nine hidden nodes.  We train the NN with 220,000 $b$ jets and test
for overtraining with 220,000 orthogonal $b$ jets. Training is done with equal amounts of leading and sub-leading jets.
The output of the NN function is a scale factor shown on Figure~\ref{nnout}, which is applied to the jet four-vector corrected with level 5 corrections. 

\section{Jet-Energy Resolution}
\label{jetreso}

\begin{figure}[H]
  \begin{center}
    \includegraphics[width=6.0cm,clip=]{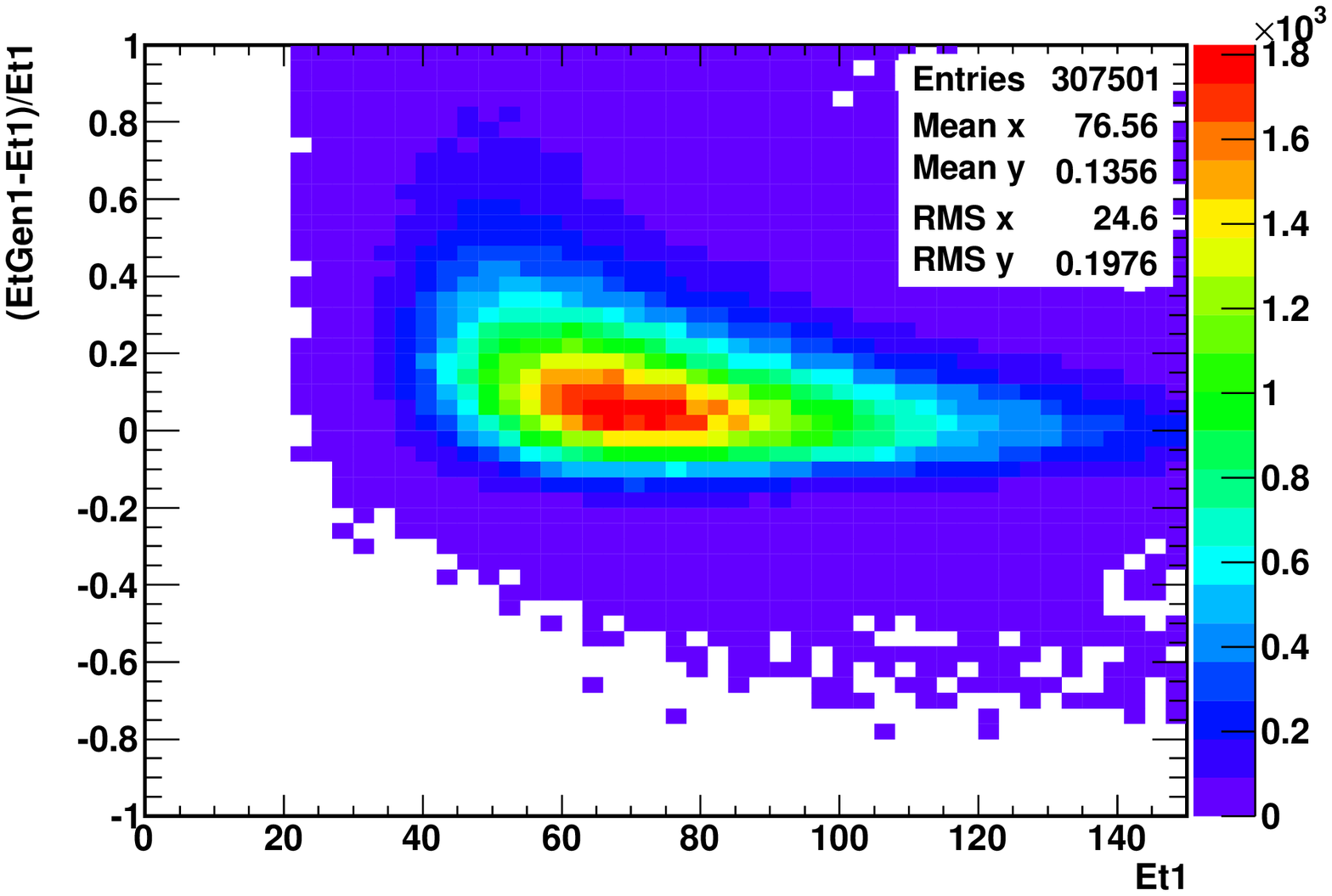}
    \includegraphics[width=6.0cm,clip=]{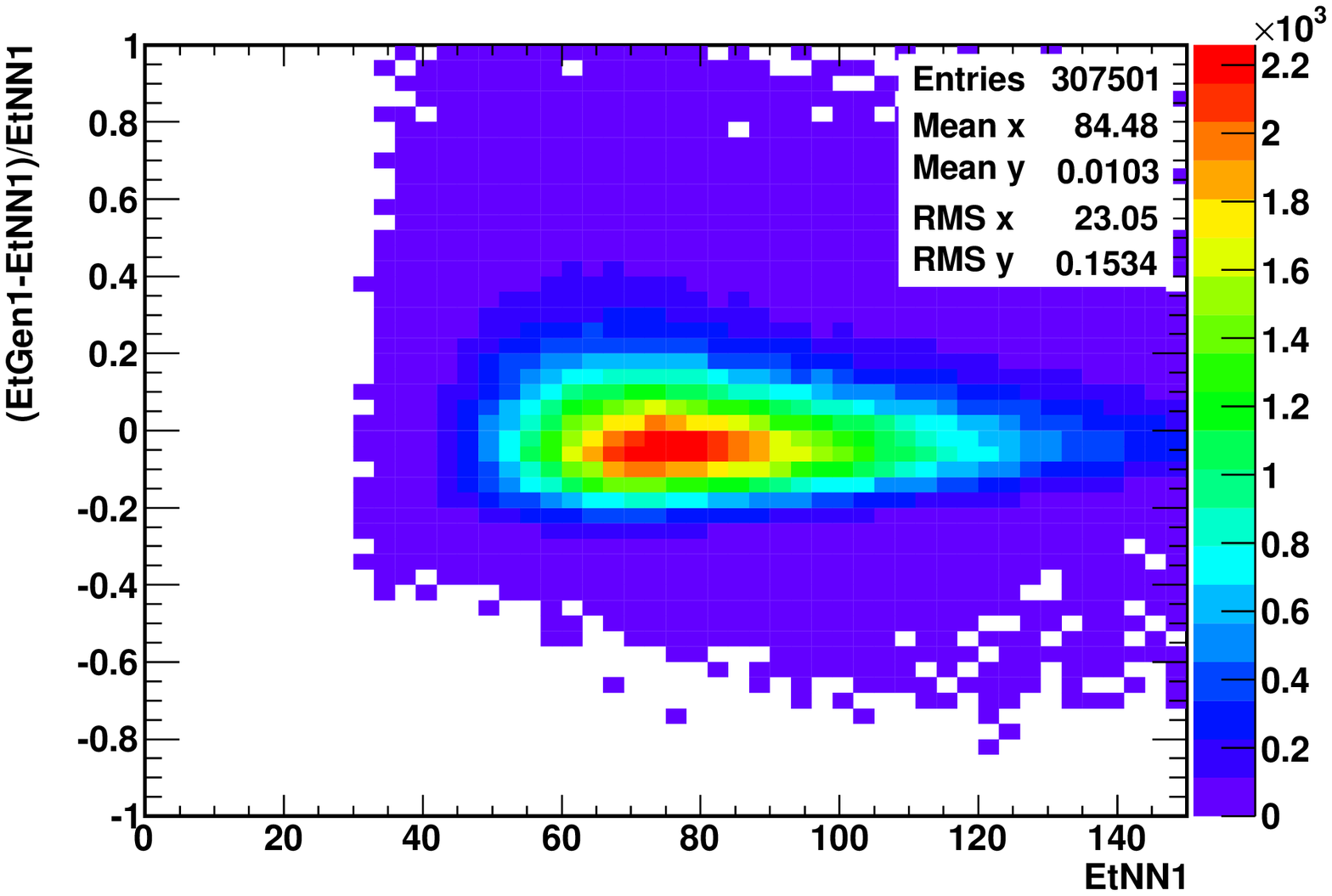}
    \caption{Comparison of the ST-tagged leading jet resolutions. The left plot
      shows the level-5 corrected jet-energy resolution as a function of the reconstructed energy and
      the right plot the NN-corrected jet-energy resolution as a function of the NN-corrected jet energy.
      Mean Y is corrected closer to the nominal values from 0.136 to 0.010 and the standard deviation
      RMS Y is corrected from 0.198 to 0.153.}
    \label{jetst1}
  \end{center}
\end{figure}

\begin{figure}[H]
  \begin{center}
    \includegraphics[width=6.0cm,clip=]{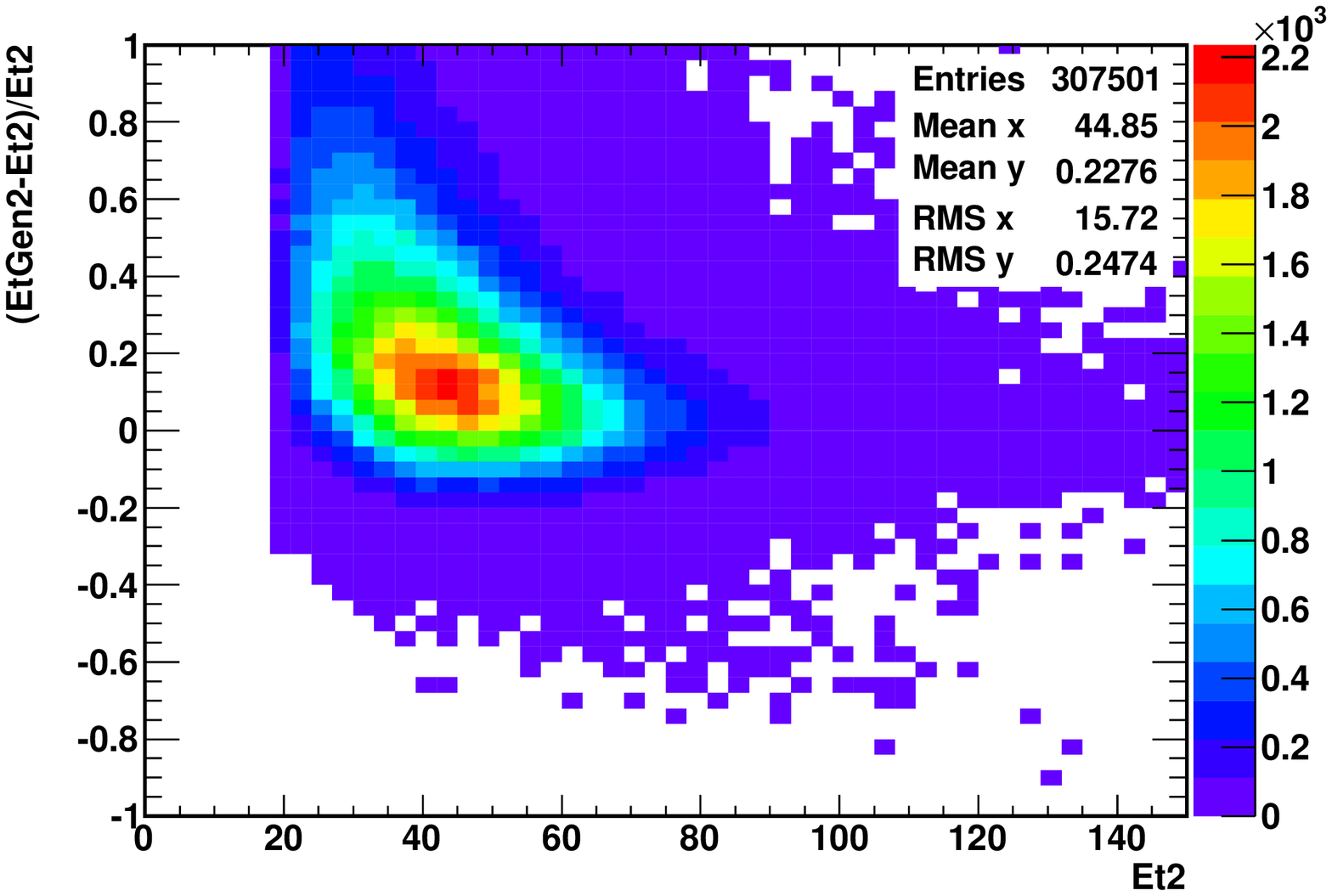}
    \includegraphics[width=6.0cm,clip=]{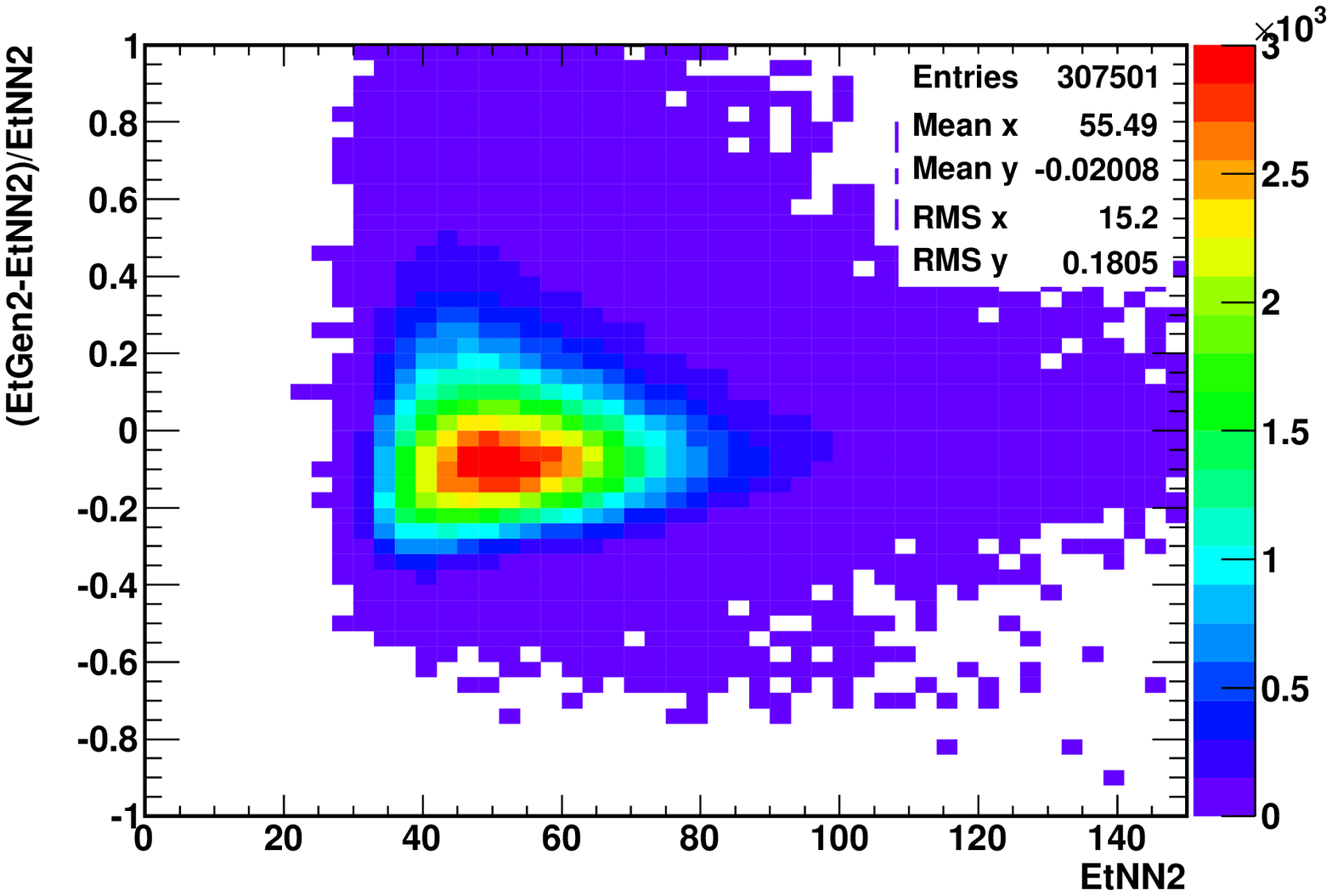}
    \caption{Comparison of the ST-tagged sub-leading jet resolutions. The left plot
      shows the level-5 corrected jet resolution as a function of the reconstructed energy and
      the right plot the NN-corrected jet resolution as a function of the NN-corrected jet energy.
      Mean Y is corrected closer to the nominal values from 0.228 to -0.020 and the standard deviation
      RMS Y is corrected from 0.247 to 0.181.}
    \label{jetst2}
  \end{center}
\end{figure}

The improvement in the jet resolution is evaluated by plotting the jet-energy resolution as a function
of the level-5 corrected jets compared to the NN-corrected jets. Leading and sub-leading
jets are checked separately. 
Figures \ref{jetst1} and \ref{jetst2} show the resolutions for the leading and
sub-leading jets separately as a function of the level 5 reconstructed energies, comparing level-5 corrected (left)
jets to those corrected by the NN jet correction function (right). We can see that both leading and sub-leading jets
are closer to their nominal values and resolution for leading jet improves from 0.198 to 0.153 and
for sub-leading jet from 0.247 to 0.181.

\section{Reconstructed Dijet Masses}
\label{recdijet}

\begin{figure}[H]
 \begin{center}
   \includegraphics[width=6.0cm,clip=]{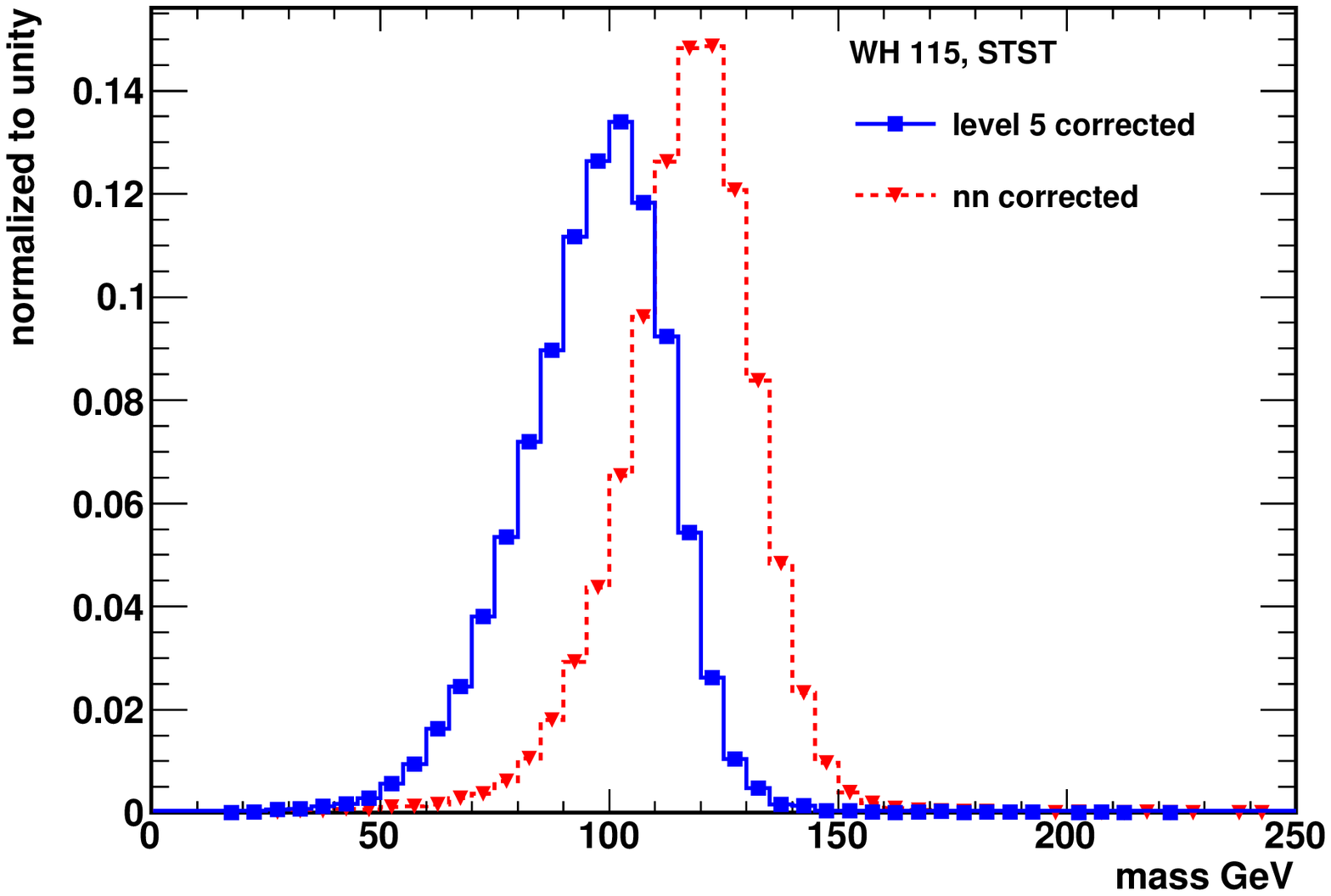}
   \includegraphics[width=6.0cm,clip=]{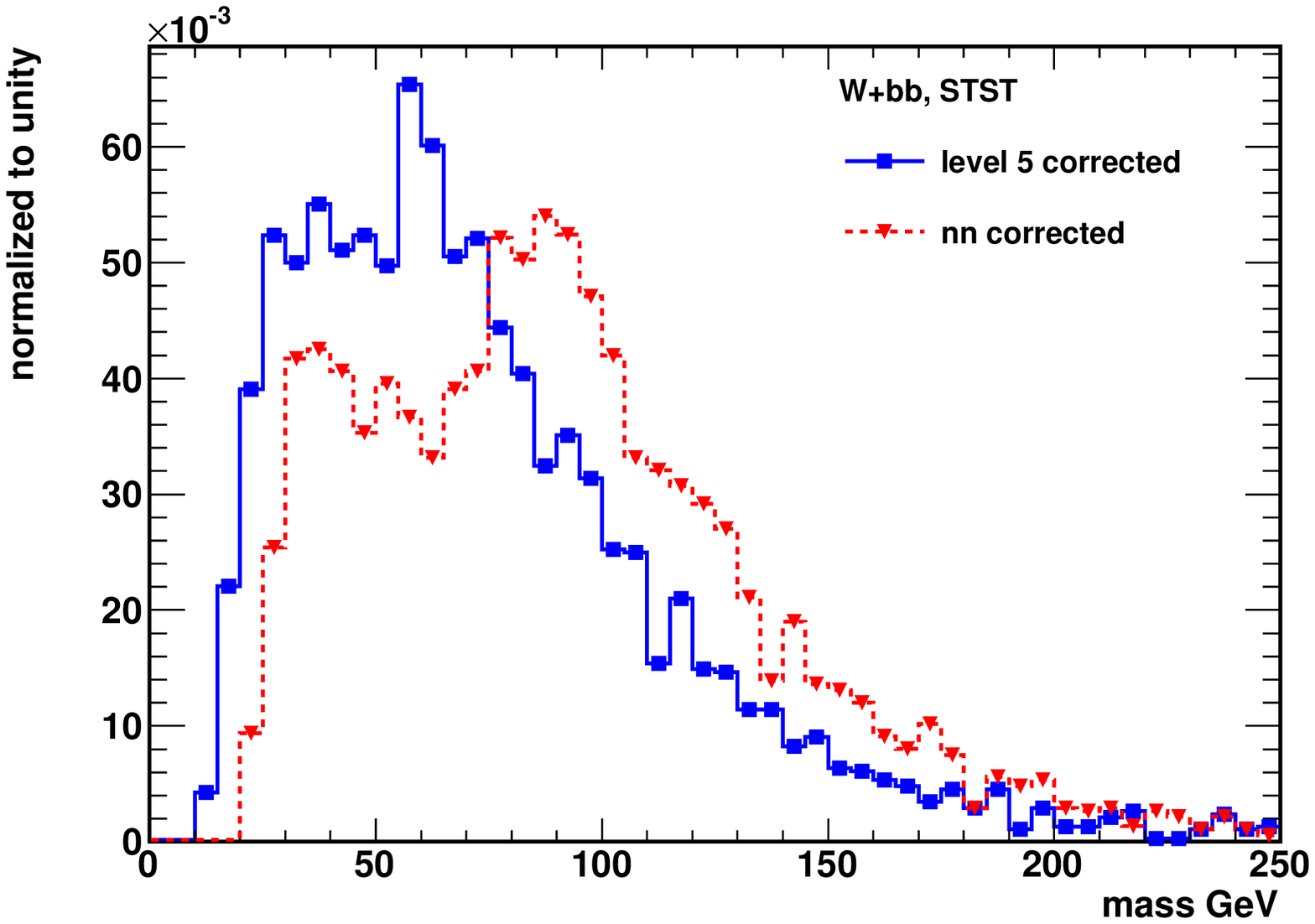}
   \includegraphics[width=6.0cm,clip=]{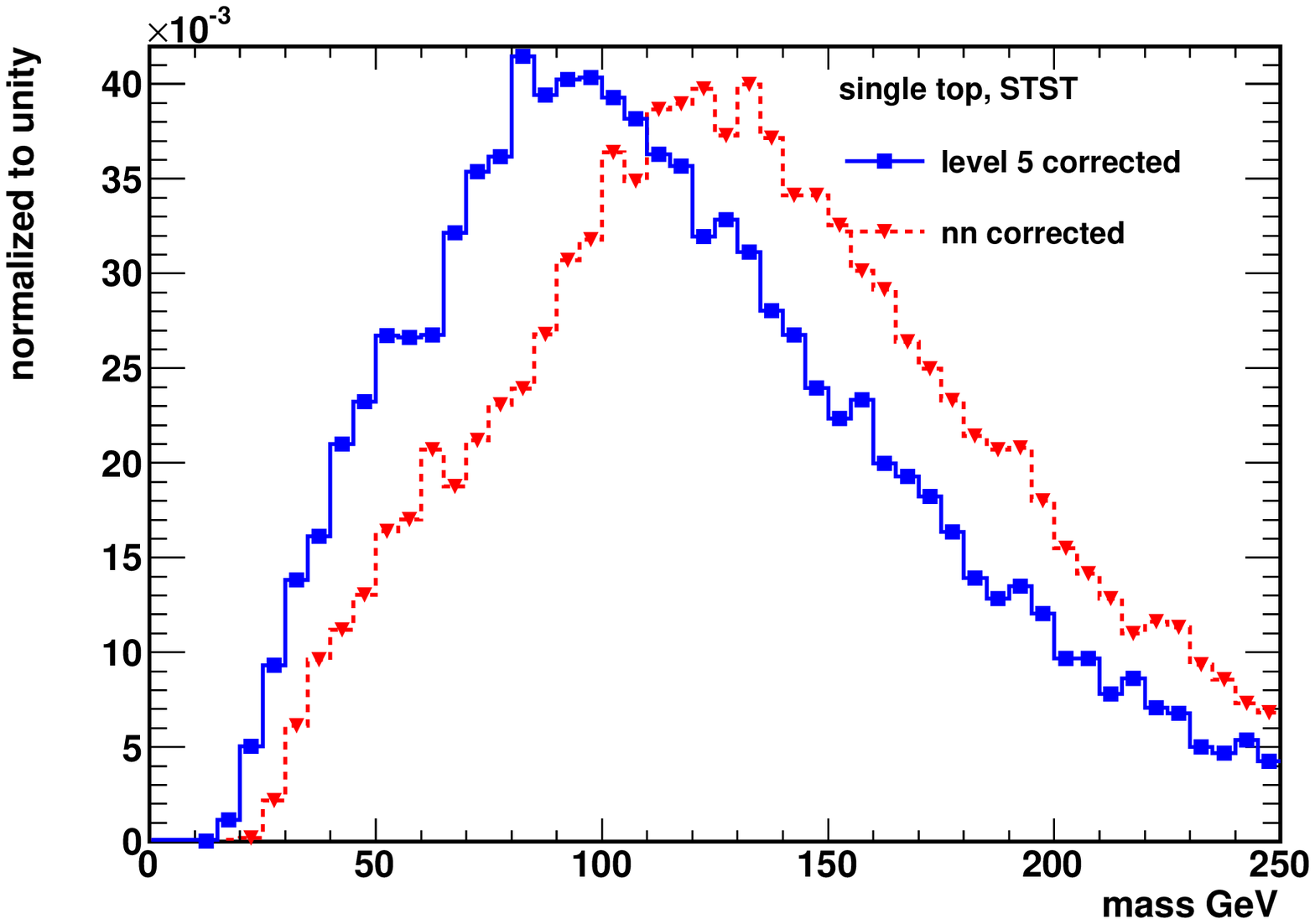}
   \includegraphics[width=6.0cm,clip=]{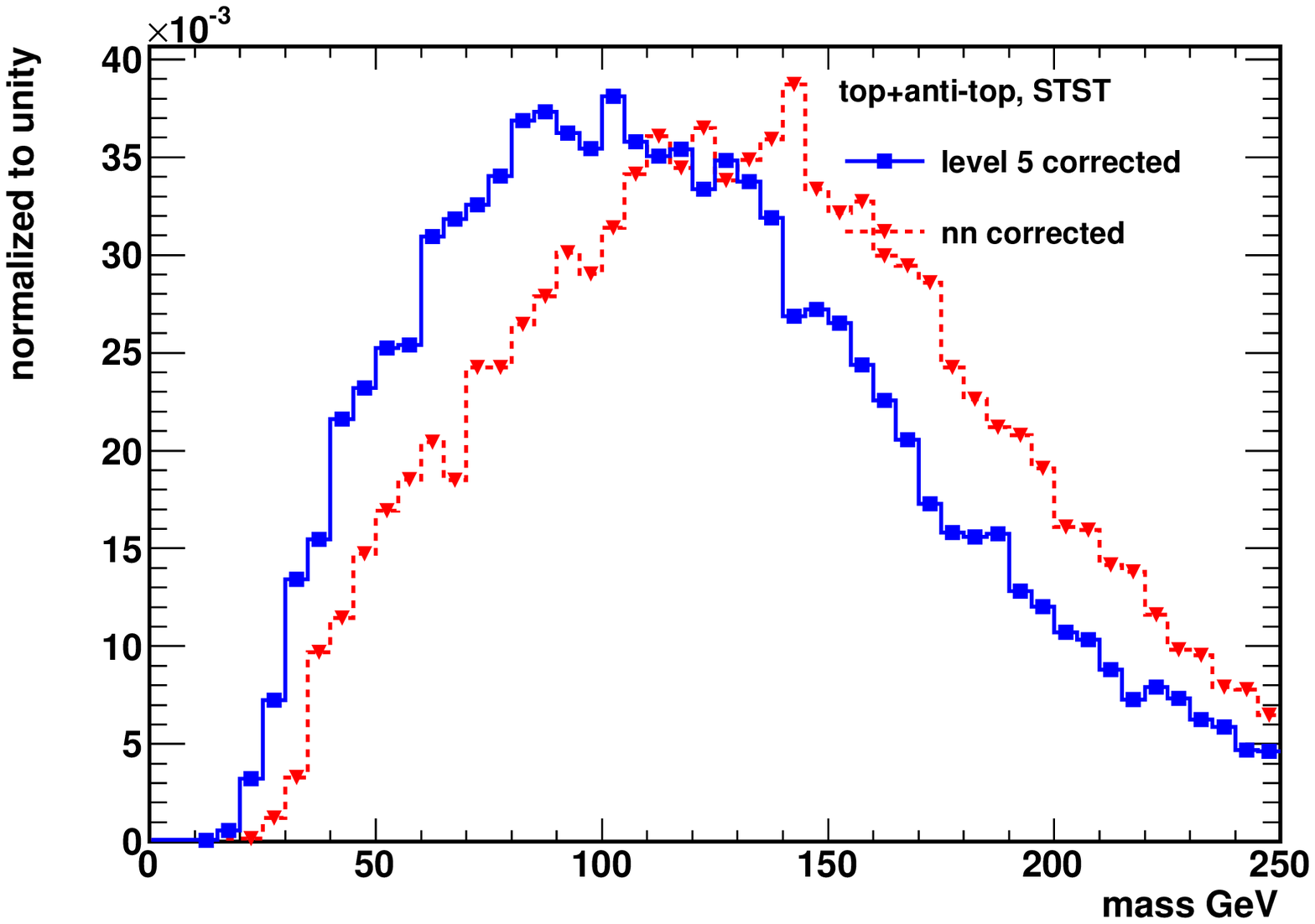}
   \includegraphics[width=6.0cm,clip=]{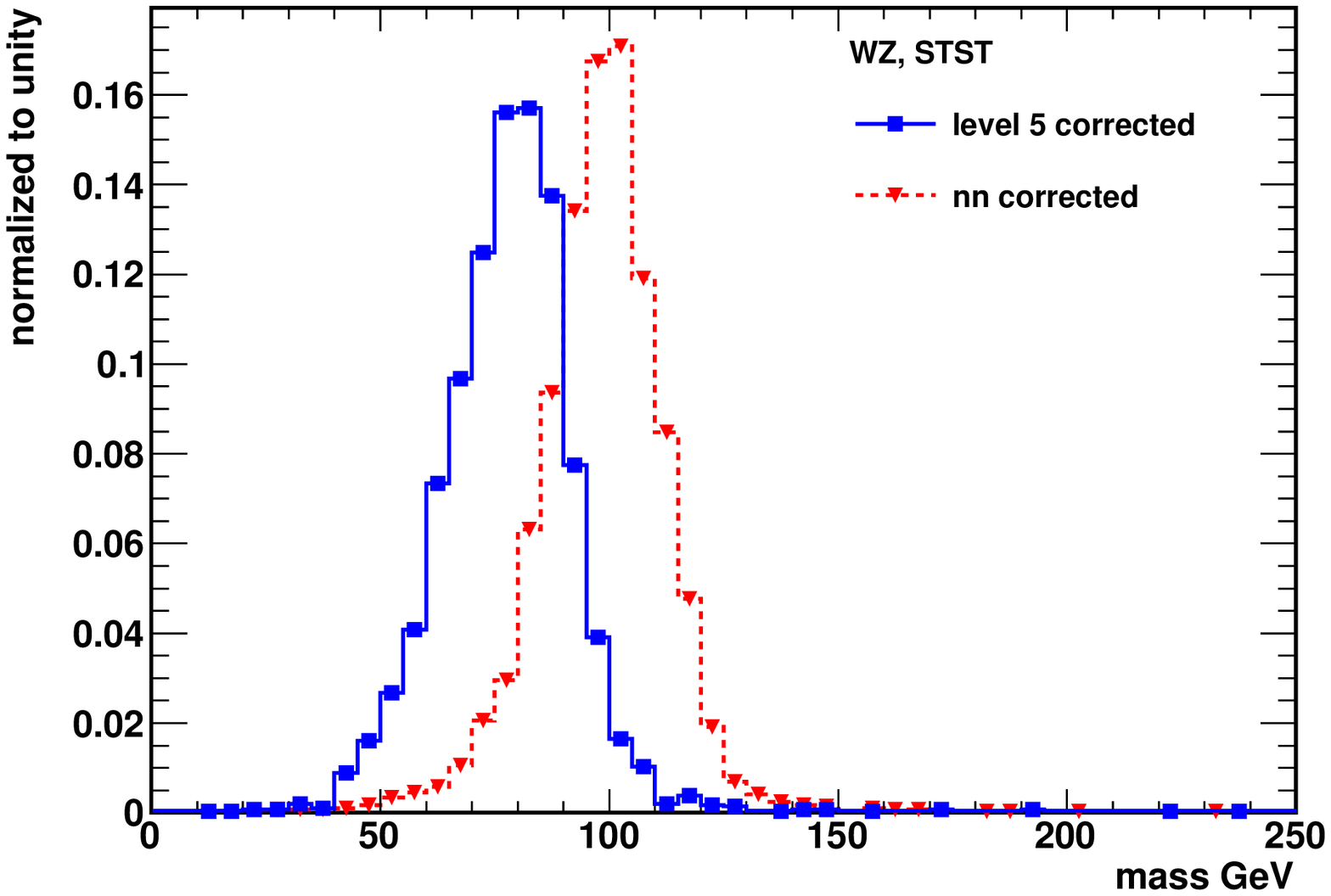}
   \caption{The comparison of the $WH$ with a generated mass of 115 GeV/$c^2$, $W+bb$, single-top, top+anti-top
     and $WZ$ dijet mass distributions with the level 5 corrections (blue solid line) and with
     the NN corrections (red dotted line) in the STST category with one central charged lepton.}
   \label{dijetmassstst}
 \end{center}
\end{figure}

After requiring two ST $b$-tags,  the main $b\bar b$ backgrounds to the $WH \rightarrow \ell \nu b\bar b$ process 
are $W+b\bar b \sim 47\%$, $t\bar t \sim 27\%$, single-top $\sim 11\%$, and QCD $\sim 8\%$. Figure \ref{dijetmassstst}
shows the reconstructed dijet mass distributions for $WH$ events in MC with a
generated mass of 115 GeV/$c^2$ and the backgrounds, excluding the QCD, which we model with data before $b$-tagging. In
addition, we show the WZ background process, which is much smaller, but interesting due to the $Z\rightarrow b\bar b$ decay.  

\section{Higgs Mass Resolution And Linearity}
\label{higgsreso}

\begin{figure}[H]
 \begin{center}
   \includegraphics[width=6.0cm,clip=]{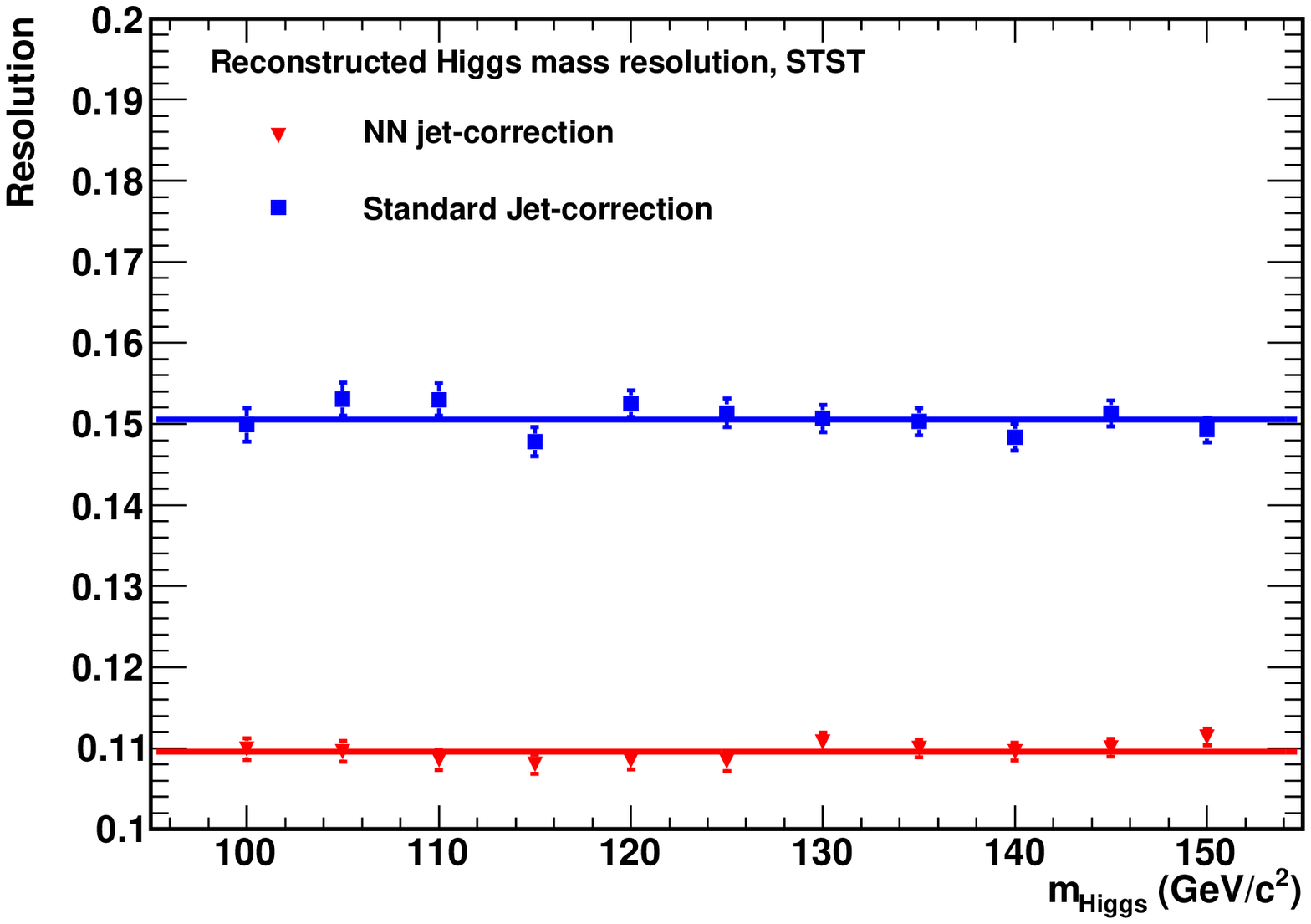}
   \includegraphics[width=6.0cm,clip=]{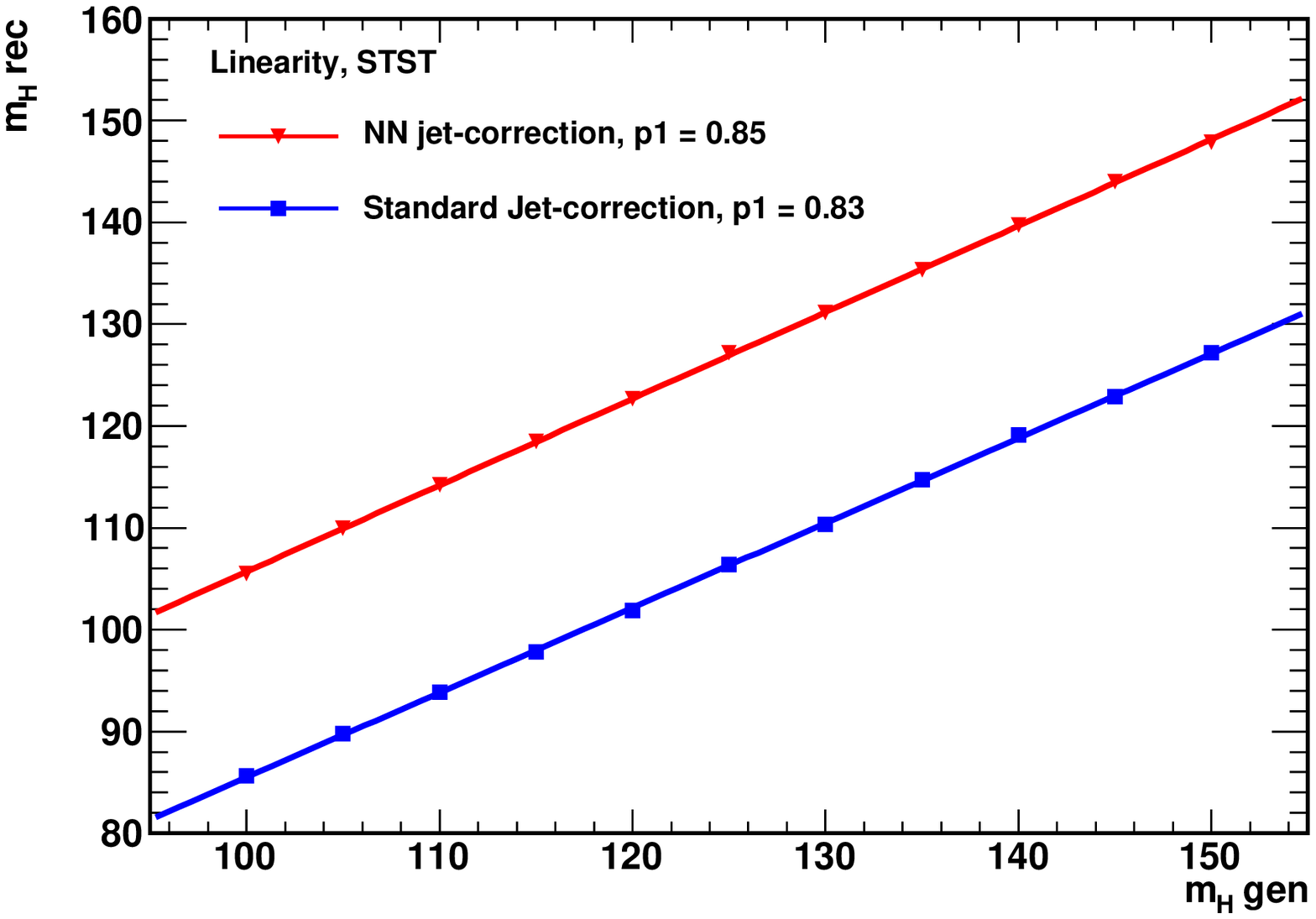}
   \caption{The left plot shows the reconstructed Higgs mass resolution in the STST category with one central charged lepton.
     The blue line shows the resolution with the level 5 corrections, with a fitted average
     value of a 15\%. The red line shows the Higgs mass resolution after applying the NN correction
     and a fitted average value of a 11\%. The right plot shows the linearity of the reconstructed
     Higgs mass as a function of generated mass.}
   \label{massres1}
 \end{center}
\end{figure}

For the $WH$ search, the main motivation to reduce the $b$-jet energy 
resolution is to improve the reconstructed $H \to b\bar{b}$ dijet invariant mass resolution, 
which appears as a resonance above a falling background continuum. The left plot on Figure \ref{massres1}
shows the improvement in the Higgs mass resolution in events with two ST tags and one central lepton, from 
the level-5 corrected jets result of $\sim$ 15\% to the NN-corrected jets result of $\sim$ 11\%,
across all the Higgs boson mass range studied.
The right plot shows that the correction is linear, such that there is no bias towards any particular 
generated mass point.

\section{Results}
\label{results}

\begin{figure}[H]
 \begin{center}
   \includegraphics[width=12.0cm,clip=]{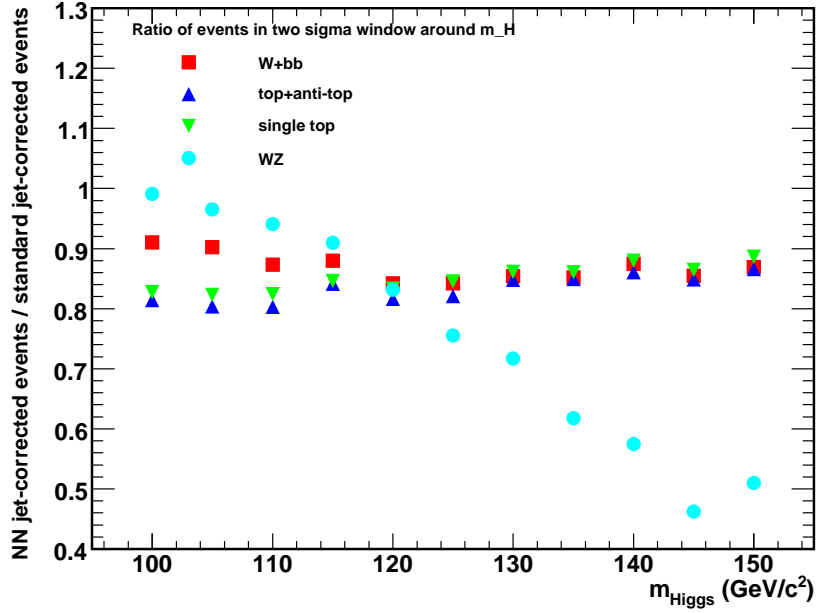}
   \caption{The plot shows the ratio of different main backgrounds around
     two sigma window of $m_H$. The rejection for the main backgrounds
     is $\sim$ 10-20\% and for the $WZ$ from 0-50\%. Statistical uncertainties of the points
     are shown by the markers.}
     \label{rejection}
 \end{center}
\end{figure}

We test how much the signal to background ratio improves against the main backgrounds
by counting the number of background events under a two standard deviation window formed by fitting the 
reconstructed Higgs mass peak to a Gaussian function, and calculating the fraction of the number
of backgorund events with NN jet corrections divided by the number of events with standard jet corrections, shown on Figure~\ref{rejection}.
For the main backgrounds, the rejection is $\sim$ 10-20\% 
while for the $WZ$ background, the improvement increases from 0 to 50\% with mass as the $Z \to b\bar{b}$ dijet 
mass resolution also improves, further separating this background contribution from the Higgs signal. 

\begin{figure}[H]
 \begin{center}
   \includegraphics[width=6.0cm,clip=]{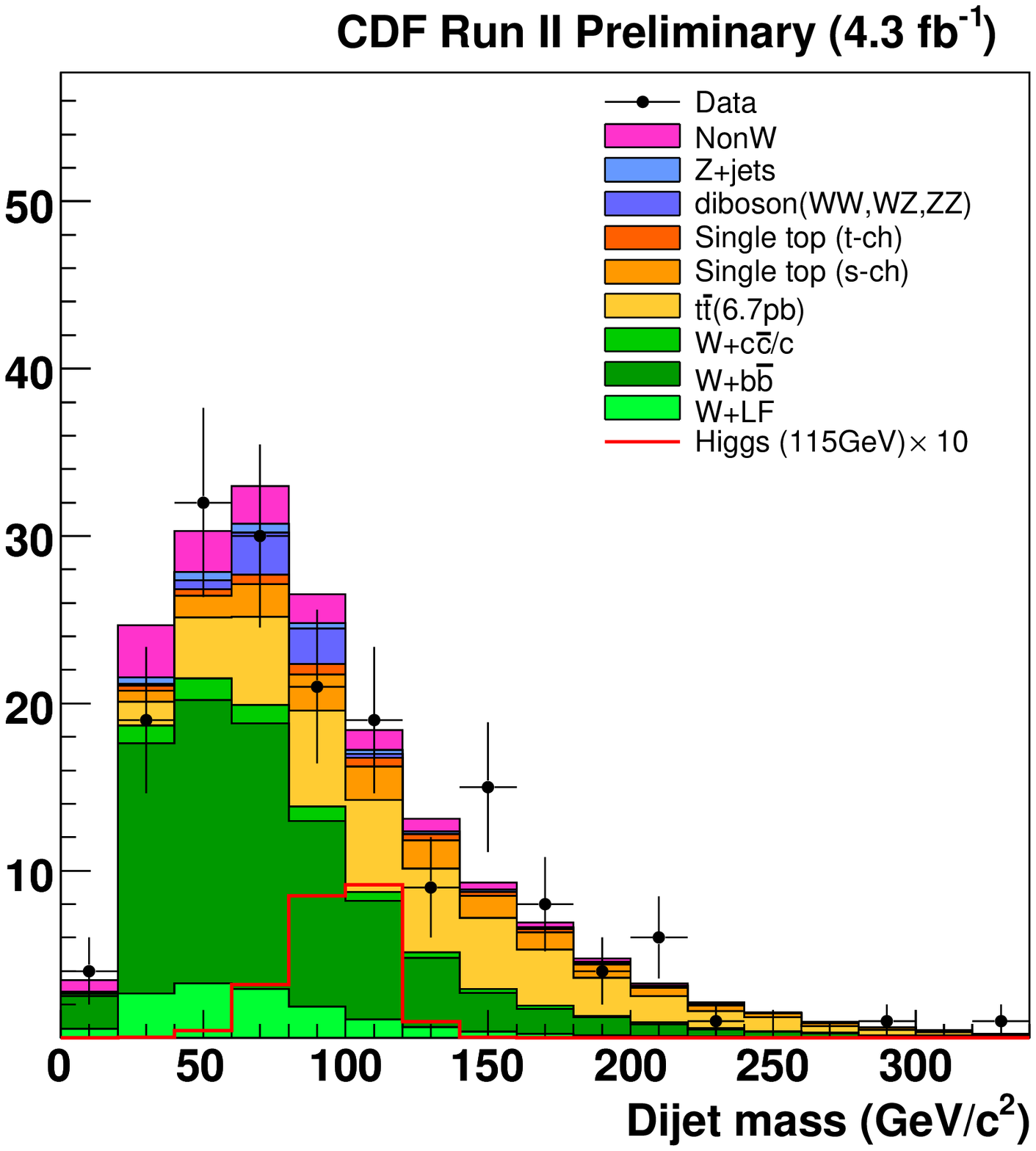}
   \includegraphics[width=6.0cm,clip=]{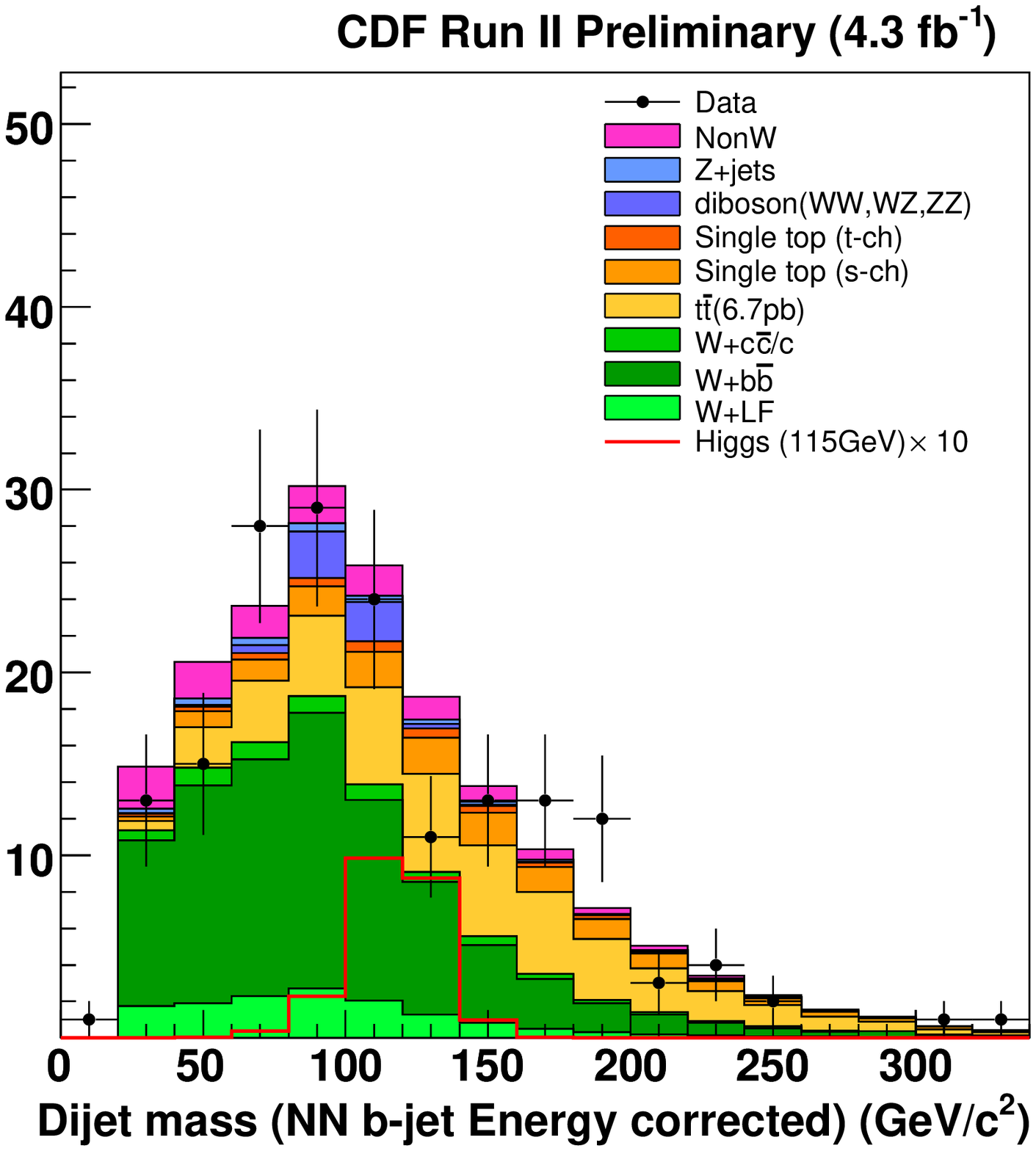}
   \caption{The reconstructed dijet mass requiring two ST-tags using only central charged leptons
     with level-5 corrected jets (left) and with NN-corrected jets (right). Using only the dijet mass
     as an input to the Higgs sensitivity calculation gives an $\sim$ 8\% improvement
     in the expected sensitivity in two tagged jets and central charged lepton region.}
     \label{dimass}
 \end{center}
\end{figure}

Using only the dijet mass shown on Figure \ref{dimass} as a discriminant
in the $WH$ analysis, we performed pseudo-experiments to test how much the dijet mass
resolution alone improves the $WH$ search sensitivity and determined 
a $\sim$ 8\% improvement in the region of two tagged jets and central charged leptons.

\begin{figure}[H]
 \begin{center}
   \includegraphics[width=6.0cm,clip=]{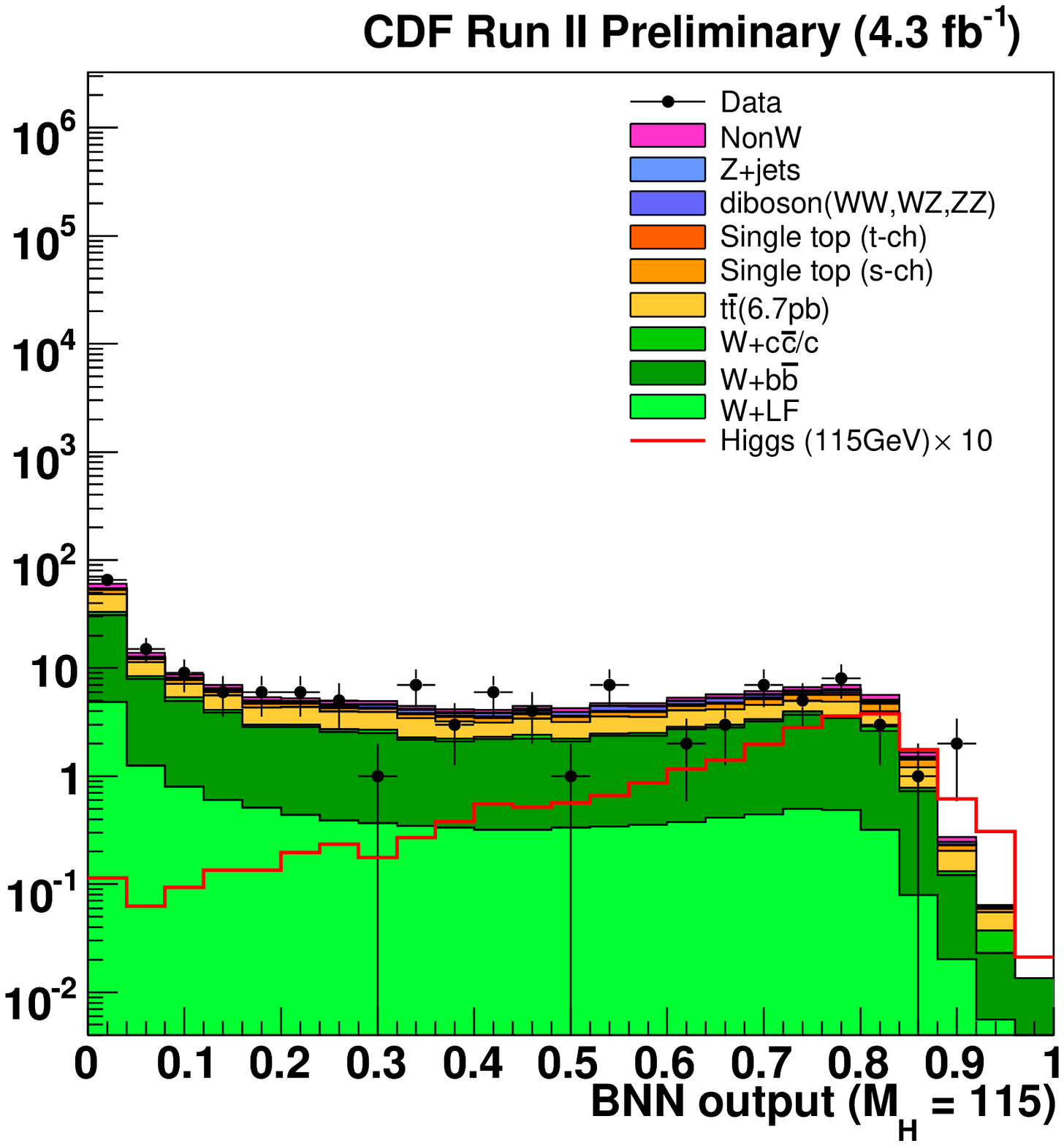}
   \includegraphics[width=6.0cm,clip=]{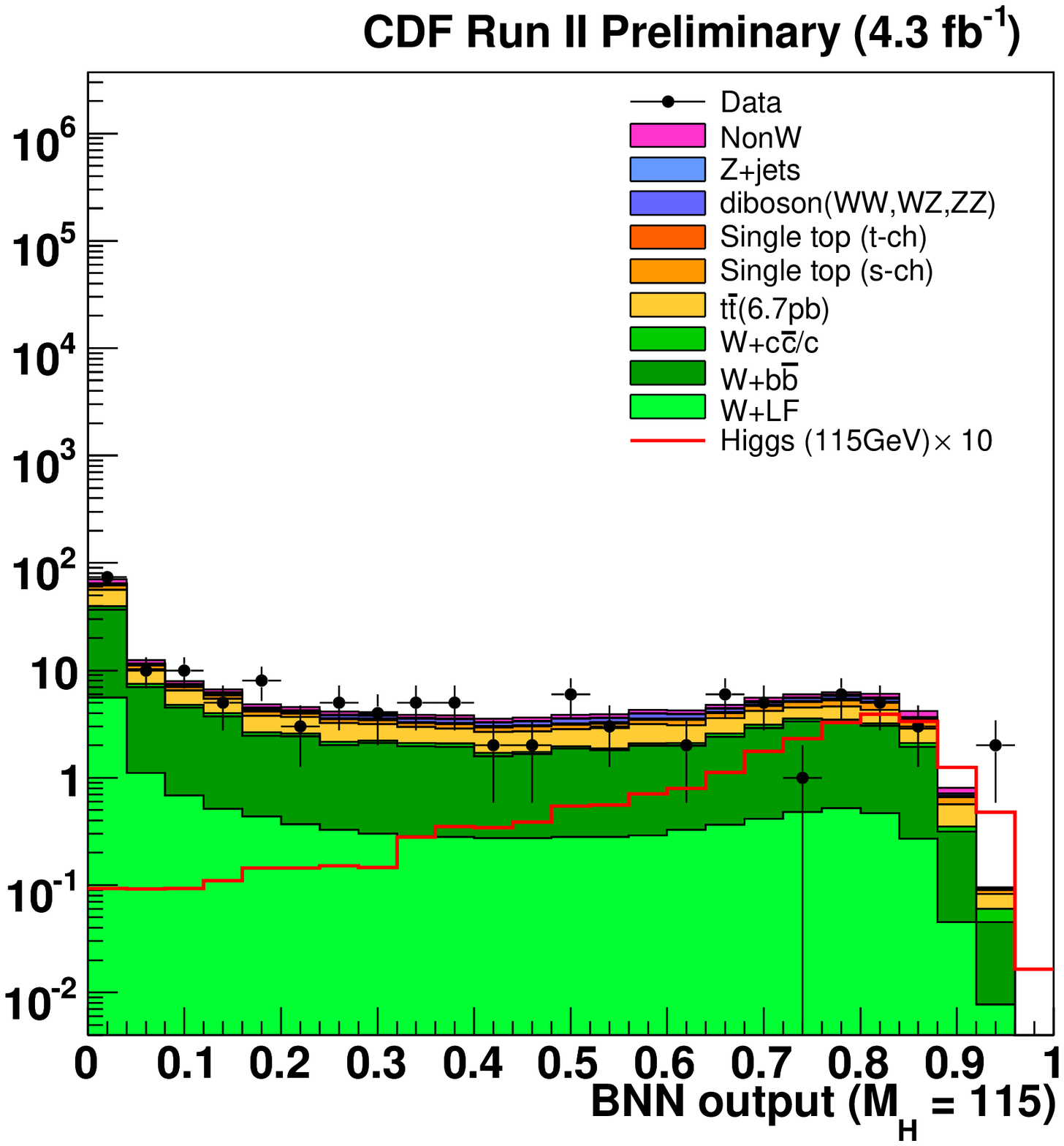}
   \caption{The Bayesian Neural Network output requiring two ST-tags and one central lepton.  The
     left plot shows the BNN output having one of the inputs the level-5 corrected dijet mass, which is
     replaced with NN-corrected dijet mass for the right plot. This translates to
     $\sim$ 9\% improvement in the expected sensitivity.}
   \label{bnnout}
 \end{center}
\end{figure}

The complete analysis used in the CDF $WH$ search uses a Bayesian Neural Network (BNN) output 
to maximize the information of the event beyond just the dijet invariant mass by taking into account the different
correlations of the kinematic distributions between the signal and background.  For instance, the signal has
a higher total event energy than the background. Combining these and some other variables
improves the separation of signal and background over a dijet mass search alone, as demonstrated by Figure \ref{bnnout}, which
shows the separation between signal and background, which the BNN output achieves. 

We train two separate BNN functions, one using
kinematic quantities calculated from standard level-5 corrected jets, and one from our NN-corrected jets. 
We find an expected improvement in the Higgs boson search sensitivity of $\sim$ 9\% in our most sensitive
search region, which is two tagged jets and one central charged lepton only.  

Besides Higgs boson searches, there are ongoing efforts at CDF to use similar corrections in measurements of the production cross section
of $WZ \to \ell \nu b\bar{b}$ and singly-produced top quarks.  Preliminary tests show that the Z-boson mass resolution improves with this
correction from 15.4\% to 11.6\% in $WZ$ events with two ST-tags and one central charged lepton.

\section{Summary}
\label{summary}

We have introduced a method to correct the $b$-jet energies using calorimetry, tracking, and vertexing information. 
We have developed a NN-regression function, which has been successfully used in a recent $WH \rightarrow \ell \nu b\bar b$ search at
the CDF. Applying the correction to the $b$ jets in $WH \rightarrow \ell \nu b\bar b$ in events with two ST-tags and one central charged lepton 
improves the dijet invariant mass resolution from $\sim$ 15\% (level-5 corrected jets) to $\sim$ 11\% (NN-corrected jets),
reduces the main $b\bar b$ backgrounds by $\sim$ 10-20\% under the two standard deviation window of Higgs mass peak,
and improves the final sensitivity of the CDF $WH$ analysis $\sim 9\%$ for events with two ST-tags and one central lepton. 



\begin{thebibliography}{00}
\bibitem{coordinates} CDF II uses a cylindrical coordinate system in which $\phi$
  is the azimuthal angle, $r$ is the radius from the nominal
  beamline, and $z$ points in the proton beam direction with
  the origin at the center of the detector. The transverse
  plane is the plane perpendicular to the $z$ axis. The $\hat x$ axis
  points outward from the accelerator ring, and the $\hat y$ axis
  points upwards.
\end{thebibliography}
\end{document}